\documentclass[a4paper,twocolumn,nofootinbib,superscriptaddress,aps,prl,,eqsecnum,notitlepage,showkeys]{revtex4-1}
\usepackage{multirow}
\usepackage{graphicx}
\usepackage{hyperref}
\usepackage{gensymb}
\usepackage{amsmath}
\usepackage{fontawesome}
\usepackage{dcolumn}    
\usepackage{ifpdf}
\usepackage{amssymb,lineno,amsfonts}
\usepackage{graphicx}   
\usepackage{bm}         
\usepackage{bbm}
\usepackage{mathrsfs}
\usepackage{upgreek}
\usepackage{mathtools}
\usepackage{epstopdf}
\usepackage{setspace}
\usepackage{hyperref}
\usepackage{natbib}
\usepackage{esvect}
\usepackage{times}
\usepackage{soul}
\usepackage[usenames,dvipsnames]{xcolor}
\definecolor{med-blue}{RGB}{25,25,112}
\hypersetup{colorlinks, linkcolor={blue},citecolor={Blue}, urlcolor={blue}}
\usepackage [english]{babel}
\usepackage [autostyle, english = american]{csquotes}
\MakeOuterQuote{"}
\begin{document}
\title{Phase coexistence and negative thermal expansion in the triple perovskite iridate Ba$_{3}$CoIr$_{2}$O$_{9}$}
\author{Charu Garg}
\affiliation{Department of Physics, Indian Institute of Science Education and Research, Dr. Homi Bhabha Road, Pune 411008, India}
\author{Antonio Cervellino}
\affiliation{Swiss Light Source, Paul Scherrer Institute, CH-5232 Villigen, Switzerland}
\author{Sunil Nair}
\affiliation{Department of Physics, Indian Institute of Science Education and Research, Dr. Homi Bhabha Road, Pune 411008, India}
\affiliation{Centre for Energy Science, Indian Institute of Science Education and Research, Dr. Homi Bhabha Road, Pune 411008, India}
\date{\today}

\begin{abstract} 
The anomalous thermal expansion in a layered 3$d$-5$d$ based triple perovskite iridate Ba$_{3}$CoIr$_{2}$O$_{9}$ is investigated using high resolution synchrotron diffraction. Below the magneto-structural transition at 107\,K, the onset of antiferromagnetic order is associated with a monoclinic distortion of the hexagonal structure. Deeper within the magnetically ordered state, a part of the monoclinic phase distorts even further, and both these structural phases co-exist down to the lowest measured temperatures. We observe negative thermal expansion in this phase co-existence regime, which appears to be intimately connected to the temperature driven relative fractions of these monoclinic phases. The significant NTE observed in this system could be driven by magnetic exchange striction, and is of relevance to a number of systems with pronounced spin orbit interactions.

\end{abstract}
\pacs{Pacs}
\maketitle

Negative Thermal Expansion (NTE) - the counter intuitive contraction of solids on heating - has been the focus of extensive experimental and theoretical investigations for many years \cite{Barrera_2005,10.3389/fchem.2018.00267,Takenaka_2012,miller2009negative}.  By virtue of their structural diversity and chemical flexibility, the transition metal oxides have proven to be a fertile playground for uncovering new candidates which exhibit this phenomena \cite{C4CS00461B,PhysRevB.78.134105,GARCIAMUNOZ1997854,PhysRevB.55.14987,PhysRevB.54.R756}. From the structural perspective, this phenomena is feasible in systems where the crystallographic framework consists of strongly bonded polyhedra units (also called rigid unit modes) \cite{dove_heine_hammonds_1995}, and here the NTE occurs as a consequence of the transverse vibrations of these units. NTE can also arise due to underlying electronic and magnetic phenomena \cite{doi:10.1088/1468-6996/15/1/015009,SAHA2011494,takenaka2017colossal}. A significant advance in this field was the observation of giant NTE in ZrW$_{2}$O$_{8}$ \cite{Mary90,ernst1998phonon}, which then paved the way for this phenomena to be discovered in a number of other transition metal oxides. A variety of structural classes like the zeolites, perovskites and its variants, as well as spinels are now known to exhibit NTE \cite{PhysRevLett.98.147203,PhysRevB.97.134117,villaescusa2001variable,woodcock1999negative,SAHA2011494}. 

The $5d$ based transition metal oxides, specially iridates, have recently gained considerable attention given the diverse electronic, magnetic and structural ground states they exhibit. A number of exotic phenomena ranging from spin-orbit liquids, charge ordering, superconductivity, Weyl semi metals, heavy fermions have now been reported \cite{doi:10.1146/annurev-conmatphys-031115-011319}. Owing to the extended nature of the iridium 5\textit{d} orbitals, several energy scales like the on-site Hubbard interaction \textit{U}, Hund's coupling J$_{H}$, the SOC $\lambda$, the crystal field $\Delta$ and the hopping energy \textit{t} become comparable; and a complex interplay between these competing scales result in their multifunctional attributes. A relatively recent addition to the strongly correlated iridates are the layered triple perovskites of the form $A_{3}X$Ir$_{2}$O$_{9}$ where $A$ is an alkaline earth metal, and $X$ can be an alkali metal, alkali earth metal, 3\textit{d} transition metal, or a lanthanide. Most of these systems adopt a hexagonal 6H triple perovskite in which the crystal structure is based on $X$O$_{6}$ octahedra connected via corners to face sharing Ir$_{2}$O$_{9}$ dimers. Given the complex geometry of corner and face sharing octahedras, the lattice is frustrated and extremely susceptible to perturbations. It has been observed that this frustration is typically released in the form of a magnetic and structural phase transformation\cite{PhysRevB.85.041201, PhysRevLett.108.217205, PhysRevB.90.014403, Lee_2016, Ranjith_2017}. In spite of the structural diversity and chemical flexibility offered by these iridates, a functionality which has not been seriously explored pertains to the feasibility of NTE in these systems. 

Here, we report structural and magnetic investigations of Ba$_{3}$CoIr$_{2}$O$_{9}$ - a 3\textit{d}-5\textit{d} based member of this triple perovskite family. This system exhibits a structural transition at 107\,K, accompanied by the onset of canted antiferromagnetic order. Unlike any other reported triple perovskite, this system shows two  structural transitions with clear evidence of phase co-existence below 70\,K.  Moreover, we observe signs of magnetic stress driven negative thermal expansion at low temperatures in both these coexisting phases. Such a correlation between the stress and the  crystallographic structure is possible due to the open framework structure and flexible network intrinsic to this system, where open spaces facilitate lattice deformations and octahedral rotations. This stress, which has a magnetic origin, could be important in a number of related systems with large spin-orbit interactions. 

Polycrystalline specimens of Ba$_{3}$CoIr$_{2}$O$_{9}$ were synthesized using the solid state synthesis route, and details of the synthesis and characterization have been reported previously \cite{garg2020evolution}. Synchrotron x-ray diffraction measurements were performed using the Materials Science (MS) X04SA beam line (wavelength 0.56526$\lambda$)  at the Swiss Light Source (SLS, PSI Switzerland\cite{SLS}. The powder sample was filled in a 0.3mm capillary and the experiments were carried out in the temperature range 4.2\,K-295\,K. The crystallographic structure was  analyzed by the Rietveld method using the FullProf refinement program~\cite{Fullprof}. The structures shown in the manuscript are drawn using Vesta~\cite{vesta}. Magnetization measurements were performed using a Quantum Design (MPMS-XL) SQUID magnetometer. 

Ba$_{3}$CoIr$_{2}$O$_{9}$ crystallizes in an aristotype 6H-type BaTiO$_{3}$ hexagonal ($P6_{3}/mmc$) structure at room temperature, consistent with the previously reported triple perovskites\cite{PhysRevB.85.041201,hexa}.  This structure constitutes of two IrO$_{6}$ octahedra sharing a face along the $c$ axis, with these Ir$_2$O$_9$ dimers being connected by corner sharing CoO$_{6}$ octahedra. The structure of the triple perovskites of the form  $A_{3}BB'_{2}O_{9}$ depends on the relative radii of $A$, $B$ and $B'$ ions \citep{IrIr}. As the size of the $A$ cation increases, the $BO_{6}$ and $B'O_{6}$ octahedra rotate in order to minimize the lattice energy. This tilting of octahedra then modifies bond lengths and bond angles, as a consequence of which the effective crystallographic symmetry is reduced.   The room temperature lattice parameters of Ba$_{3}$CoIr$_{2}$O$_{9}$ as obtained from Reitvled refinement of x-ray diffraction data are a = b = 5.7639$\AA$, c = 14.2949$\AA$ and $\alpha$=$\beta$= 90$\degree$, $\gamma$= 120$\degree$. On lowering temperature, the structure transforms from the hexagonal $P6_{3}/mmc$ to a monoclinic $C2/c$ symmetry at the magneto-structural transition temperature of 107\,K, with this symmetry being retained down to 80\,K. On further lowering of the temperature, the system partially transforms to an even lower symmetry ($P2/c$), with both these structurally disparate phases coexisting down to the lowest measured temperatures.
\begin{figure}
\centering
	\includegraphics[scale=0.34]{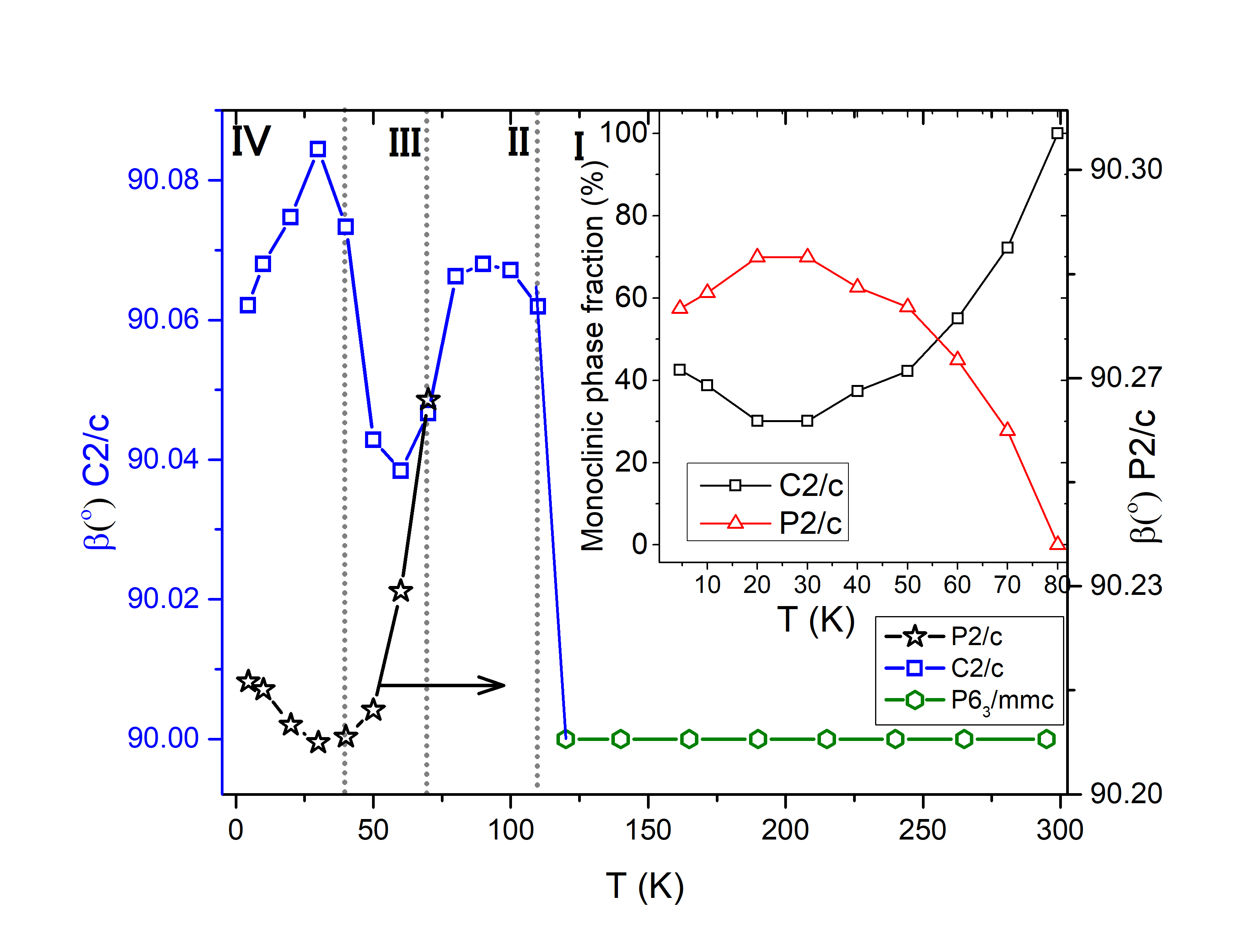}
  \caption{Main panel: Temperature dependence of the monoclinic angle $\beta$ for the P6$_{3}$/mmc ($\bigcirc$), C2/c ($\square$) and P2/c (\faStarO) phases of Ba$_{3}$CoIr$_{2}$O$_{9}$. The variation of phase fraction of the monoclinic $P2/c$ and $C2/c$ phases with temperature is depicted in the inset. }
	\label{crystal structure}
\end{figure} 

The evolution of relative ratios of both the monoclinic phases with temperature is shown in the inset of Fig1. For all the crystallographic parameters deduced from analysis of diffraction data, the error bars are of the same order (or smaller) than the symbol size used. As the temperature is reduced, the $C2/c$ phase gradually converts to the lower $P2/c$ symmetry. Interestingly, close to 40\,K,(T$_p$) this phase ratio is reversed and a part of the system slowly transforms back to the higher $C2/c$ monoclinic symmetry. The evolution of these phases can be clearly observed in $\beta$ which is the angle between the crystallographic $a$ and $c$ axes, as depicted in the main panel of Fig1. The plot can be divided into 4 regions. Region I corresponds to the (paramagnetic) hexagonal phase where $\beta$ remains constant based on symmetry constraints. In region II, there is a sharp increase in $\beta$ because of the change in symmetry at the magnetostructural transition. Region III corresponds to the phase coexistence regime, and as the $C2/c$ phase transforms to $P2/c$, the $\beta$ value sharply increases till the phase cross over temperature (T$_p$) is reached. As soon as a maximum tilt angle of $90.084\degree$ is achieved, the phase reversal is triggered. This corresponds to region IV, where $\beta$ decreases with a concomitant increase in the relative fraction of the $C2/c$ phase. A similar trend is also observed for the lower symmetry $P2/c$ phase, with temperature dependence of $\beta$ and the phase fraction being inversely correlated to each other.
\begin{figure}
\centering
	\includegraphics[scale=0.32]{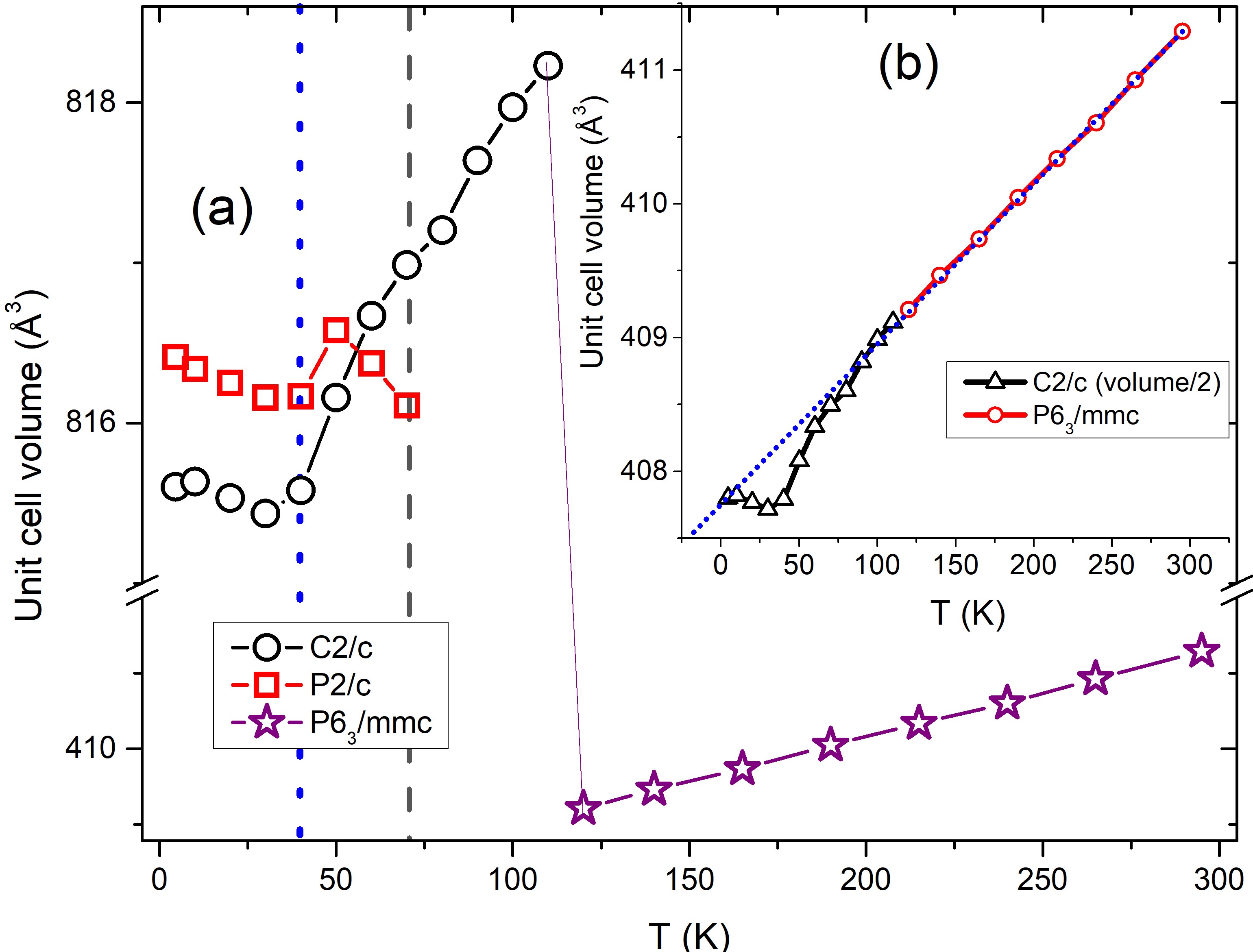}
  \caption{(a) Temperature dependence of volume as determined for P6$_{3}$/mmc (\faStarO)), $C2/c$ ($\bigcirc$) and $P2/c$ ($\square$). The vertical bar at 70K and 40K indicate the onset of phase co-existence and the phase cross over temperature T$_{P}$ respectively. (b) The reduced volume for $C2/c$ is plotted for comparison and a strong upturn is observed close to 40\,K. (c) Variation of the coefficient of volume thermal expansion $\alpha_V$  for the $C2/c$ and $P2/c$ phases.}
	\label{crystal structure}
\end{figure} 

On careful observation, these regions are also clearly demarcated in the temperature dependence of the volume, as is shown in Fig2(a). A steep rise in the volume at 107\,K marks the onset of AFM order and a magnetostructural transition as confirmed by synchrotron and neutron diffraction. A dashed vertical bar at 70\,K indicates the onset of second phase transition concomitant with the onset of a phase co-existence region. Interestingly, the monoclinic $P2/c$ phase shows an anomalous behavior where the volume increases as the temperature is lowered. This negative thermal expansion is observed right from the phase transformation at 70\,K with a discontinuity at the phase crossover temperature T$_p$ of 40\,K. Since the magnetostructural transition from the hexagonal $P6_{3}/mmc$ to the monoclinic $C2/c$ phase at 107\,K is accompanied by unit cell doubling, the reduced volume of the $P6_{3}/mmc$ phase and the volume of the $C2/c$ phase are plotted against temperature in Fig2(b) fo an easy comparison. A subtle change in the slope the structural transition at 107\,K is followed by an marked upturn in the volume at 40\,K as is also depicted in the main panel. This clearly indicates the presence of negative thermal expansion in the $C2/c$ phase as well below the phase crossover temperature T$_p$.  

The crossover from positive to negative thermal expansion results from the balance between two competitive factors: anharmonic phonon vibration and magnetic exchange-striction, with each contributing to positive and negative thermal expansion respectively. The typical value of the  coefficient of volume thermal expansion $\alpha_V$ [ = (1/V)$\textit{d}$V/$\textit{d}$T] for the prototypical NTE system ZrW$_{2}$O$_{8}$ is -2.7 x 10$^{-5}$ K$^{-1}$. The highest values of $\alpha_V$ for the $P2/c$ and $C2/c$ phases of Ba$_{3}$CoIr$_{2}$O$_{9}$ are  -1.21 x 10$^{-5}$ and -0.8 x 10$^{-5}$ K$^{-1}$ respectively, which is of the same order as that in reported in many prototypical NTE materials \cite{doi:10.1063/1.2147726,doi:10.1002/anie.201102228}. 
\begin{figure}
\centering
	\includegraphics[scale=0.29]{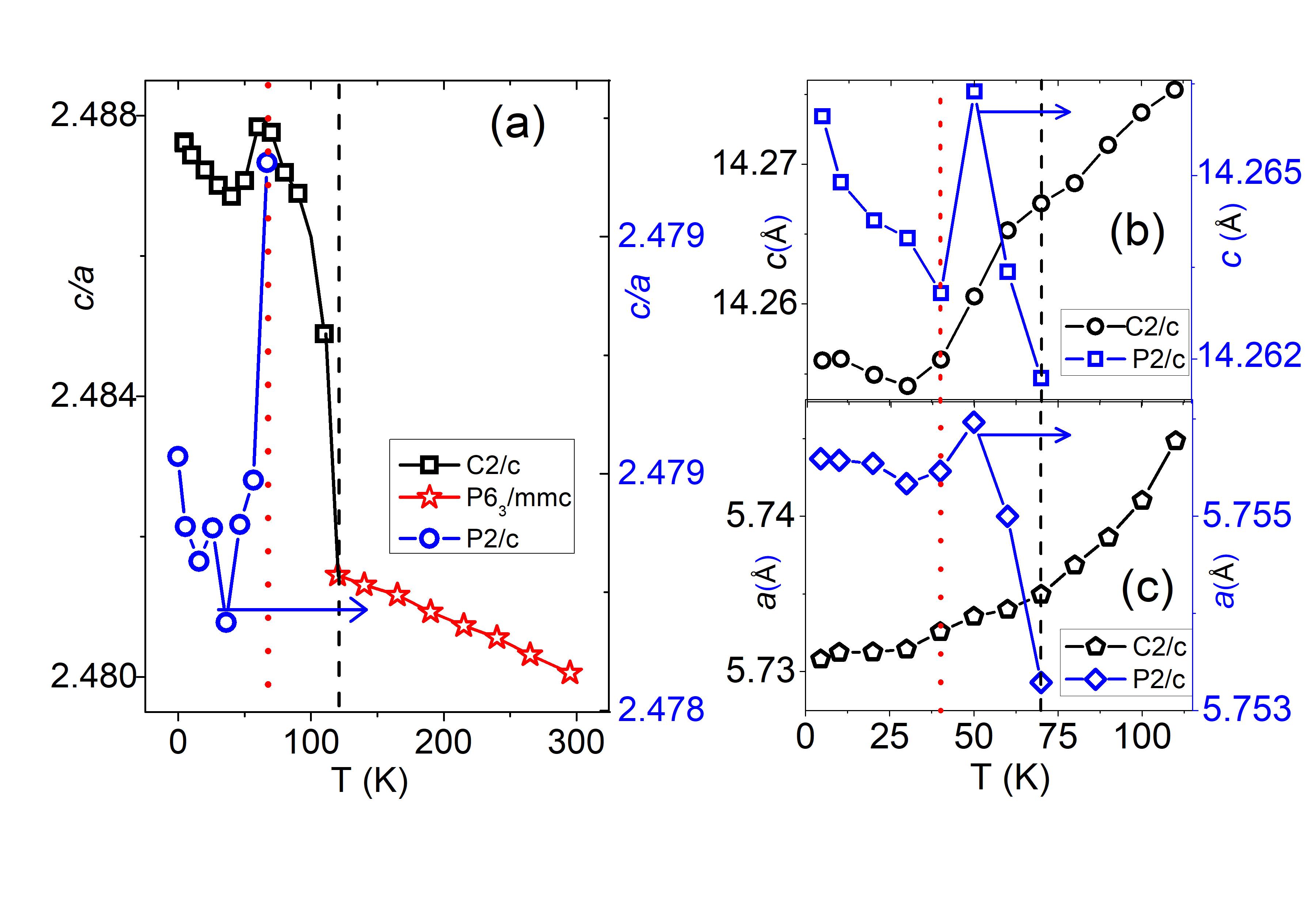}
  \caption{(a)  Temperature dependence of the $c/a$ ratio for all the structural phases encountered in Ba$_{3}$CoIr$_{2}$O$_{9}$. The dashed bar indicates the onset of magneto-structural transition and the dotted line represents the start of phase co-existence regime. Variation of lattice parameters $c$ and $a$ with temperature is depicted in (b) and (c) respective. The dashed and dotted line indicates onset of phase co-existence and the phase cross-over temperature respectively. }
	\label{crystal structure}
\end{figure}

The temperature dependence of the $c/a$ ratio for all the structural phases is depicted in Fig3(a), and significant change is observed just below the magnetostructural transition. This is indicative of a structural distortion arising due to magnetic exchange striction - which in antiferromagnetic transition metal oxides results in the generation of a strain field aimed at increasing the magnetic exchange energy \cite{PhysRevLett.85.5166}. The lattice parameter $c$ for both the monoclinic phases is depicted in Fig3(b).  For the $P2/c$ phase, it increases as a function of decreasing temperature for the entire temperature range, with a discontinuity at  T$_p$.  On the other hand, in the $C2/c$ phase, a negative slope is attained only at  T$_p$. Interestingly, the lattice parameter $a$ shows a NTE in the $P2/c$ phase and a Positive Thermal Expansion (PTE) in the $C2/c$ phase throughout the temperature range as shown in Fig3(c). The contrasting temperature dependence of the lattice parameters is indicative of anisotropic thermal expansion and also reinforces the critical influence which the lattice parameter $c$ has on the NTE in this material.
 
The fact that none of the other triple perovskites reported till date are known to exhibit NTE suggests that this phenomena is driven by the phase coexistence and phase reversal which is unique to the Ba$_{3}$CoIr$_{2}$O$_{9}$ system. There are broadly two types of NTE \cite{10.3389/fchem.2018.00267}: one which arises solely due to a structural origin - often termed as conventional NTE. In these systems, the structural framework consists of strongly bonded polyhedra units called RUMs (rigid unit modes) \cite{dove_heine_hammonds_1995} and  NTE occurs as a consequence of the transverse vibrations of these units. Due to its structural origin, they are generally present over a broad temperature range. The other type of NTE is driven by phase transitions, with underlying phenomena like charge transfer \cite{azuma2011colossal}, ferroelectric order \cite{chen2013effectively}, magnetic and orbital ordering \cite{PhysRevB.85.165143}, giant magnetocaloric effect \cite{10.3389/fchem.2018.00438} being responsible. Interestingly, many oxides like LaMnO$_3$ \cite{GARCIAMUNOZ1997854,PhysRevB.55.14987}, SrRuO$_3$  \cite{PhysRevB.54.R756}, AlFeO$_3$  \cite{SAHA2011494} exhibit this type of NTE. 
\begin{figure}
\centering
	\includegraphics[scale=0.5]{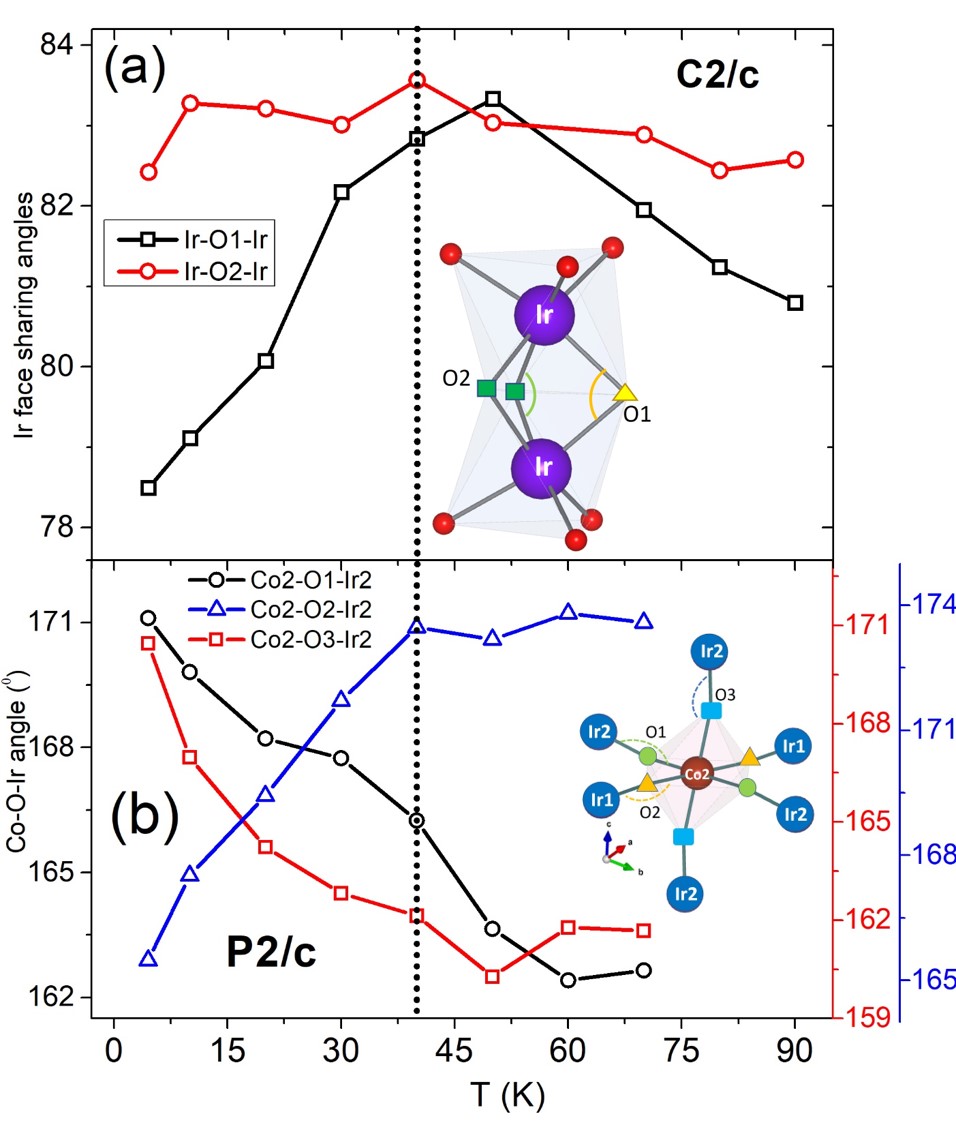}
  \caption{(a) Main panel: Thermal evolution of face sharing Ir-O-Ir angles within the Ir dimer depicted pictorially in the inset. O1($\blacktriangle$) and O2($\blacksquare$) represents the face sharing oxygen. (b) Main panel: Temperature variation of the 3 unique Co-O-Ir angles made by cobalt and Iridium in the $C2/c$ and $P2/c$ phases. Inset: A schematic representation of the cobalt octahedra connected via corner to Iridium. The vertical dotted line represents the phase crossover temperature T$_{P}$.}
	\label{crystal structure}
\end{figure}
Prerequisites for observing negative thermal expansion by the transverse thermal motion of oxygen include: (i) an open framework structure with oxygen in a two-fold coordination, (ii) strong metal-oxygen bonds to avoid significant thermal expansion of the bond length, and (iii) rigid polyhedra which are free to tilt back and forth with little or no change in shape. The monoclinically distorted triple pervskite structure of Ba$_{3}$CoIr$_{2}$O$_{9}$ appears to be susceptible to distortions allowed by the RUM model due to the flexible network and open framework structure intrinsic to this symmetry. Interestingly, these rotations of the IrO$_{6}$ and CoO$_{6}$ octahedra also allows for the overall monoclinic symmetry to be preserved while transforming to a lower space group. Fig4(a) displays the change in the Ir-O-Ir bond angles in the face sharing dimer for the $C2/c$ phase. Although the Ir-O2-Ir angle remains relatively rigid below the magnetostructural transition, the Ir-O1-Ir angle has a strong temperature dependence. As soon as it reaches the maxima of octahedral tilting at the phase cross over temperature (T$_{P}$), the octahedra are buckled, as is reflected in the sharp decrease in the Ir-O1-Ir angle.  A similar trend is seen in Co-O-Ir for the $P2/c$ monoclinic phases as shown in the Fig4(b), where a distinct change of slope corresponding to the buckling of the octahedra is observed. 
\begin{figure}
\centering
	\includegraphics[scale=0.5]{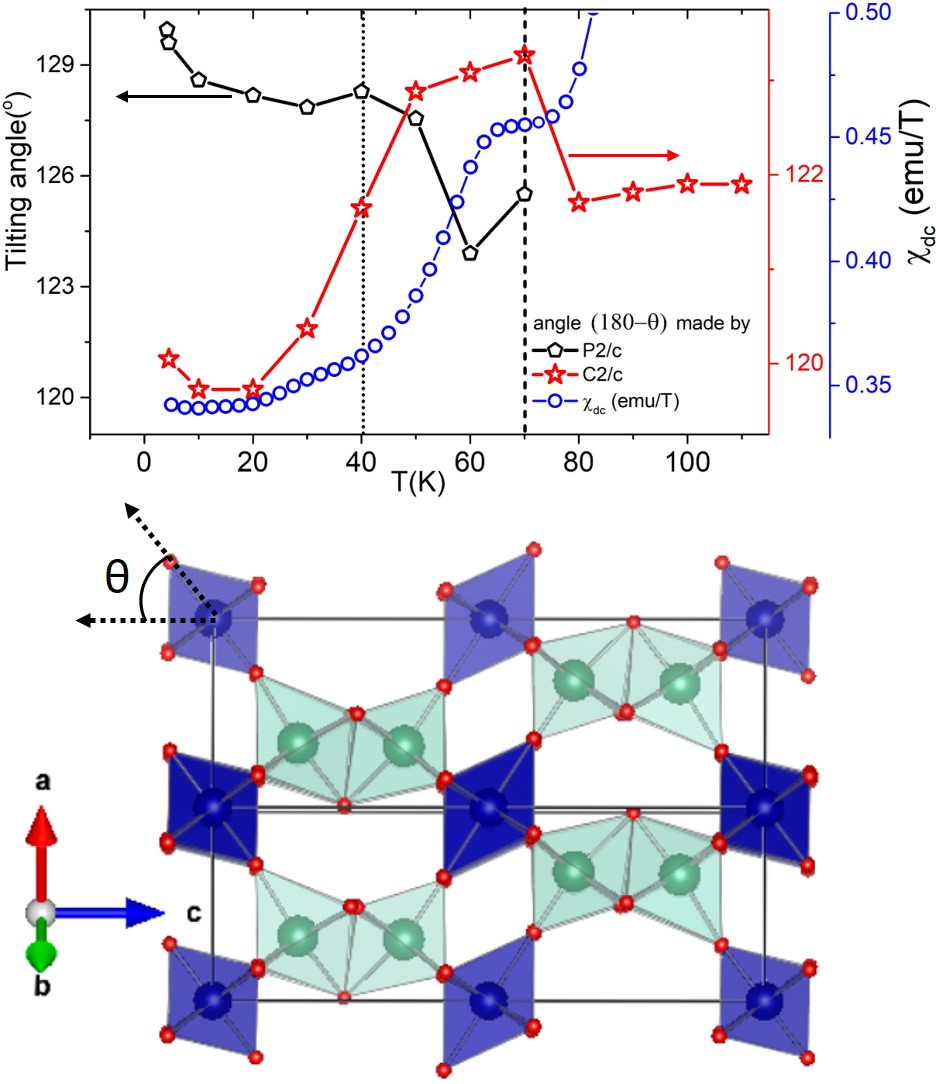}
  \caption{Top panel: Temperature-dependent dc magnetization data measured in a magnetic field of 100 Oe for Ba$_{3}$CoIr$_{2}$O$_{9}$. Red and black curves indicates the tilt angle of CO$_{6}$ octahedra for C2/c and P2/c respectively, as shown in the schematic below. The CO$_{6}$ octahedra sits at the corner and $\theta$ is the angle formed by the apical oxygen with the c-axis.}
	\label{crystal structure}
\end{figure}

This correlation between the NTE and the octahedral distortions is seen even more clearly in the temperature dependence of the tilt of the CoO$_6$ octahedra which sit at the corners of the unit cells in both the $C2/c$ and the $P2/c$ phases, as is depicted in Fig5(a). In the mononclinic $C2/c$ phase, this angle is initially invariant, and increases on approaching the phase co-existence regime. The stabilization of the mononclinic $P2/c$ phase partially relieves this rotation, and this quantity is seen to be inversely correlated in both the phases. This is an indicator of how the lattice strain in one of the phases is relieved by the stabilization of the other. Moreover, signatures of this phase coexistence is also seen as a hump in the dc magnetization, which is also plotted in the same graph for clarity. Though there is no evidence of a low temperature magnetic transition in this system, it is evident that the magnetization is modulated by the crystallographic structure,  indicating the intimate connection between the two. 

Our prior neutron diffraction and first principles calculations indicate that the spins are predominantly aligned along the crystallographic $c$-axis in this system \cite{garg2020evolution}. With the structural distortions and magnetization appearing to be intimately coupled, it is not surprising that the major lattice change is observed along this direction. Fig6 depicts the hysteresis in the magnetization as measured in the Field Cooled Cooling (FCC) and the Field Cooled Warming (FCW) protocols in the phase coexistence regime. This indicates the first order nature of this transition, which is associated with formation of a new phase $P2/c$ inside the monoclinic $C2/c$ phase which is stabilized below the magnetostructural transition at 107\,K.

\begin{figure}
\centering
	\includegraphics[scale=0.3]{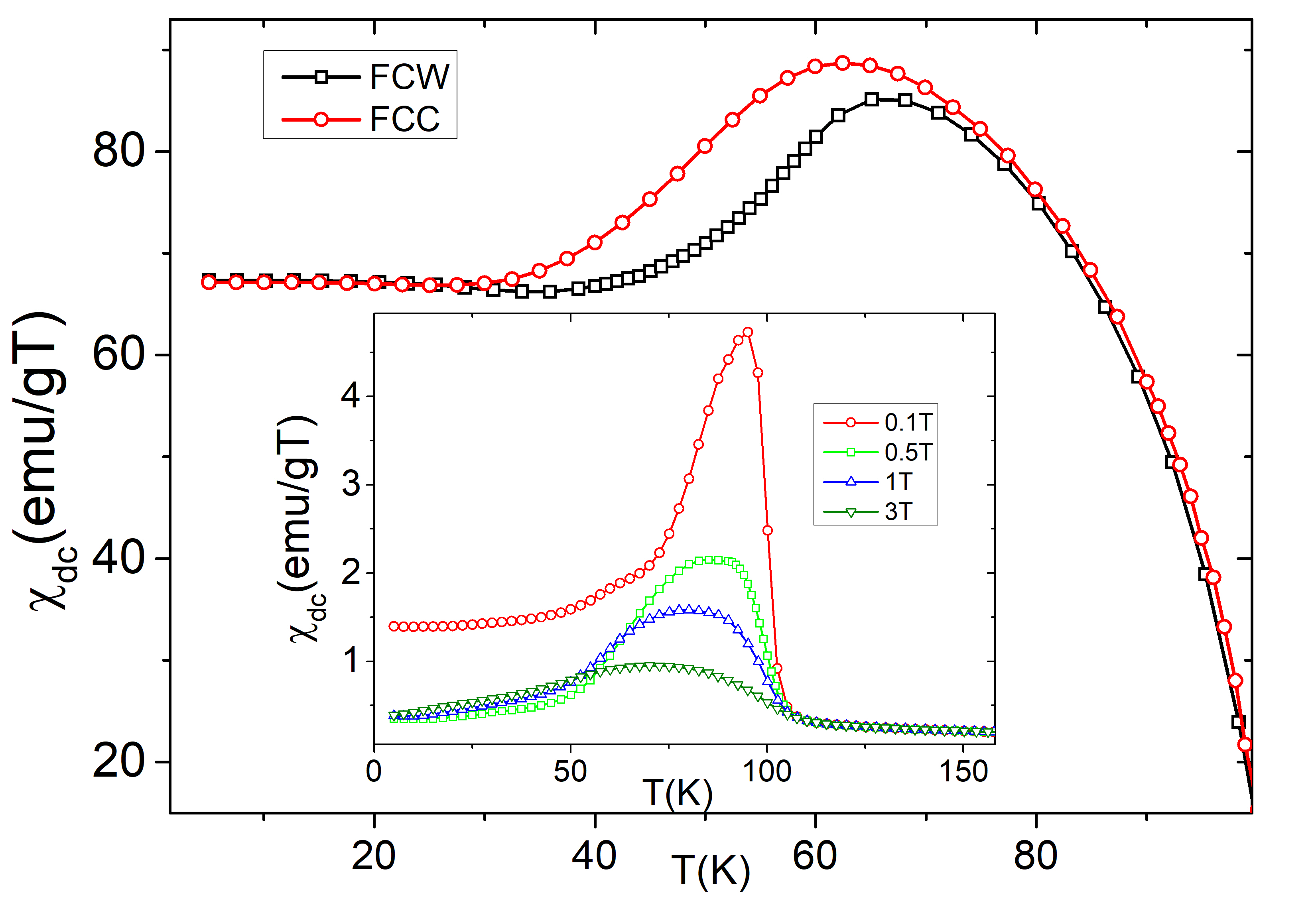}
  \caption{Main panel:) Temperature-dependent dc magnetization data measured under a magnetic field of 100Oe under field-cooled warming (FCW) and field-cooled cooling (FCC) protocols. Inset: Temperature dependence of zero field-cooled (ZFC) dc magnetization data of Ba$_{3}$CoIr$_{2}$O$_{9}$ at different magnetic fields.}
	\label{crystal structure}
\end{figure}
This intimate coupling between the magnetism and NTE is reminiscent of the large magnetostriction and NTE observed in the frustrated spinel ZnCrSe$_{4}$ \cite{PhysRevLett.98.147203}. It was suggested that the NTE in that system arose as a consequence of a highly frustrated lattice, with exchange-striction being the driving mechanism. We note that in  Ba$_{3}$CoIr$_{2}$O$_{9}$ which orders magnetically at 107\,K, a fit to the high temperature magnetization data gave a Curie-Weiss temperature ($\theta_{\rm CW}$) of 6.35\,K, which is clearly suggestive of mixed ferro-antiferromagnetic interactions. This was also confirmed by measurements of MH isotherm, where a small loop opening -typical of ferri-, or weak ferromagnetic systems- was observed. The magnetic structure in the $C2/c$ phase was deduced to be a canted antiferromagnet, with the main antiferromagnetic component along the $c$ axis. Though the antiferromagnetic magnetic structure in the $P2/c$ phase remains to be ascertained, it is likely that it only differs in the extent of the canting angle. An interesting aspect of Ba$_{3}$CoIr$_{2}$O$_{9}$ pertains to the temperature regime in which NTE is observed. For instance, in systems like  Ca$_{2}$RuO$_{4}$\cite{takenaka2017colossal}, Pb(Ti,V)O$_{3}$\cite{pan2017colossal}, Mn$_{3}$ZnN  \cite{doi:10.1063/1.3540604}, the NTE is driven by an electronic or magnetic phase transition, and is observed (or has its onset) in the transition region. Conversely, in the ZnCrSe$_{4}$ system, NTE was observed predominantly in the paramagnetic regime (starting from $T> 3 \times T{_N}$) down to the magnetic transition temperature, owing to short range spin fluctuations characteristic to geometrically frustrated magnets. In contrast to these cases, the NTE in Ba$_{3}$CoIr$_{2}$O$_{9}$ is observed deep within the AFM state, and is intimately related to a phase coexistence between the two disparate monoclinic phases. 

This situation is similar to that reported earlier in the rare earth vanadates of the form $R$VO${_3}$ (with $R$ = Sm, Ho, Yb, Pr or Y) \cite{184107,195102}, where competing orbitally ordered phases coexist as a consequence of the competition between the lattice strain arising out of octahedral tilting and that due to exchange-striction. We speculate that a similar mechanism is at play here, and that the temperature evolution of the structurally disparate monoclinic phases in Ba$_{3}$CoIr$_{2}$O$_{9}$ arises as a consequence of the interplay between exchange striction, relative energy difference between the $C2/c$ and $P2/c$ phases, as well as octahedral rotations in each phase. On traversing the magnetostructural transition from the high temperature paramagnetic ($P6_{3}/mmc$) state, we first encounter the $C2/c$ phase which appears to be energetically most favorable. However, increasing exchange striction as a function of reducing temperature strains the lattice, which in turn is partially relieved by a conversion of a part of the $C2/c$ phase to the lower symmetry $P2/c$ phase starting from 70\,K. The relative energy difference between these monoclinic phases appear to be dynamic, and below the crossover temperature, a part of the $P2/c$ phase converts back to the higher symmetry $C2/c$ phase. However, this alone does not compensate for the exchange striction induced strain, and a part of it is accommodated by the octahedral rotations in both these disparate phases which then manifests itself in the form of NTE in \emph{both} of them. 

This interplay between competing phases is also likely to make the ground state amenable to external perturbations. A pronounced magnetic field dependence of the NTE has been previously observed in ZnCrSe$_{4}$ \cite{PhysRevLett.98.147203} where the balance between the AFM and FM states in the frustrated magnetic spinels can be easily tuned with the application of a magnetic field \cite{PhysRevLett.98.147203, PhysRevLett.97.087204}. Though both the co-existing phases in the case of Ba$_{3}$CoIr$_{2}$O$_{9}$ are AFM, it is quite likely that perturbations like pressure, or magnetic fields can tilt the balance between them. Our Zero field cooled measurements at different applied magnetic fields as shown in the inset of Fig6 indicates that the broad hump-signifying the region of phase co-existence gets suppressed and shifts to even low temperature values as a function of increasing magnetic fields. Magnetic field dependent measurements of x-ray diffraction and magnetostriction are likely to throw more light on the evolution and dynamics of these co-existing phases.
 
In summary, we report on the existence of a low temperature regime of negative thermal expansion in a new 3\textit{d}-5\textit{d} based triple perovskite iridate  Ba$_{3}$CoIr$_{2}$O$_{9}$. This unusual temperature dependence of the volume is intimately related to the co-existence of two closely related monoclinic phases, and the temperature dependence of their relative phase fractions. The underlying driving mechanism appears to be the antiferromagnetic exchange striction, which precipitates a phase conversion from a monoclinic $C2/c$ phase to a more distorted $P2/c$ phase as a function of decreasing temperatures. Below a phase crossover temperature of 40\,K, the phase ratios of these monoclinic phases inverts, with remainder of the strain being accommodated by octahedral distortions which  manifests itself in the form of a negative thermal expansion in both these phases. This work opens up opportunities in exploring NTE in intrinsically layered systems which are characterized by strong spin orbit interactions.   

S.N. acknowledges DST India for support through the DST Nanomission Thematic Unit Program, SR/NM/TP-13/2016. C.G. and S.N. thank the Department of Science and Technology, India (SR/NM/Z-07/2015) for  financial support and Jawaharlal Nehru Centre for Advanced Scientific Research (JNCASR) for managing the project. 

\bibliography{Bibli}

\begin{thebibliography}{43}%
\makeatletter
\providecommand \@ifxundefined [1]{%
 \@ifx{#1\undefined}
}%
\providecommand \@ifnum [1]{%
 \ifnum #1\expandafter \@firstoftwo
 \else \expandafter \@secondoftwo
 \fi
}%
\providecommand \@ifx [1]{%
 \ifx #1\expandafter \@firstoftwo
 \else \expandafter \@secondoftwo
 \fi
}%
\providecommand \natexlab [1]{#1}%
\providecommand \enquote  [1]{``#1''}%
\providecommand \bibnamefont  [1]{#1}%
\providecommand \bibfnamefont [1]{#1}%
\providecommand \citenamefont [1]{#1}%
\providecommand \href@noop [0]{\@secondoftwo}%
\providecommand \href [0]{\begingroup \@sanitize@url \@href}%
\providecommand \@href[1]{\@@startlink{#1}\@@href}%
\providecommand \@@href[1]{\endgroup#1\@@endlink}%
\providecommand \@sanitize@url [0]{\catcode `\\12\catcode `\$12\catcode
  `\&12\catcode `\#12\catcode `\^12\catcode `\_12\catcode `\%12\relax}%
\providecommand \@@startlink[1]{}%
\providecommand \@@endlink[0]{}%
\providecommand \url  [0]{\begingroup\@sanitize@url \@url }%
\providecommand \@url [1]{\endgroup\@href {#1}{\urlprefix }}%
\providecommand \urlprefix  [0]{URL }%
\providecommand \Eprint [0]{\href }%
\providecommand \doibase [0]{http://dx.doi.org/}%
\providecommand \selectlanguage [0]{\@gobble}%
\providecommand \bibinfo  [0]{\@secondoftwo}%
\providecommand \bibfield  [0]{\@secondoftwo}%
\providecommand \translation [1]{[#1]}%
\providecommand \BibitemOpen [0]{}%
\providecommand \bibitemStop [0]{}%
\providecommand \bibitemNoStop [0]{.\EOS\space}%
\providecommand \EOS [0]{\spacefactor3000\relax}%
\providecommand \BibitemShut  [1]{\csname bibitem#1\endcsname}%
\let\auto@bib@innerbib\@empty
\bibitem [{\citenamefont {Barrera}\ \emph {et~al.}(2005)\citenamefont
  {Barrera}, \citenamefont {Bruno}, \citenamefont {Barron},\ and\ \citenamefont
  {Allan}}]{Barrera_2005}%
  \BibitemOpen
  \bibfield  {author} {\bibinfo {author} {\bibfnamefont {G.~D.}\ \bibnamefont
  {Barrera}}, \bibinfo {author} {\bibfnamefont {J.~A.~O.}\ \bibnamefont
  {Bruno}}, \bibinfo {author} {\bibfnamefont {T.~H.~K.}\ \bibnamefont
  {Barron}}, \ and\ \bibinfo {author} {\bibfnamefont {N.~L.}\ \bibnamefont
  {Allan}},\ }\href {\doibase 10.1088/0953-8984/17/4/r03} {\bibfield  {journal}
  {\bibinfo  {journal} {Journal of Physics: Condensed Matter}\ }\textbf
  {\bibinfo {volume} {17}},\ \bibinfo {pages} {R217} (\bibinfo {year}
  {2005})}\BibitemShut {NoStop}%
\bibitem [{\citenamefont {Takenaka}(2018)}]{10.3389/fchem.2018.00267}%
  \BibitemOpen
  \bibfield  {author} {\bibinfo {author} {\bibfnamefont {K.}~\bibnamefont
  {Takenaka}},\ }\href {\doibase 10.3389/fchem.2018.00267} {\bibfield
  {journal} {\bibinfo  {journal} {Frontiers in Chemistry}\ }\textbf {\bibinfo
  {volume} {6}},\ \bibinfo {pages} {267} (\bibinfo {year} {2018})}\BibitemShut
  {NoStop}%
\bibitem [{\citenamefont {Takenaka}(2012)}]{Takenaka_2012}%
  \BibitemOpen
  \bibfield  {author} {\bibinfo {author} {\bibfnamefont {K.}~\bibnamefont
  {Takenaka}},\ }\href {\doibase 10.1088/1468-6996/13/1/013001} {\bibfield
  {journal} {\bibinfo  {journal} {Science and Technology of Advanced
  Materials}\ }\textbf {\bibinfo {volume} {13}},\ \bibinfo {pages} {013001}
  (\bibinfo {year} {2012})}\BibitemShut {NoStop}%
\bibitem [{\citenamefont {Miller}\ \emph {et~al.}(2009)\citenamefont {Miller},
  \citenamefont {Smith}, \citenamefont {Mackenzie},\ and\ \citenamefont
  {Evans}}]{miller2009negative}%
  \BibitemOpen
  \bibfield  {author} {\bibinfo {author} {\bibfnamefont {W.}~\bibnamefont
  {Miller}}, \bibinfo {author} {\bibfnamefont {C.}~\bibnamefont {Smith}},
  \bibinfo {author} {\bibfnamefont {D.}~\bibnamefont {Mackenzie}}, \ and\
  \bibinfo {author} {\bibfnamefont {K.}~\bibnamefont {Evans}},\ }\href
  {\doibase 10.1007/s10853-009-3692-4} {\bibfield  {journal} {\bibinfo
  {journal} {Journal of materials science}\ }\textbf {\bibinfo {volume} {44}},\
  \bibinfo {pages} {5441} (\bibinfo {year} {2009})}\BibitemShut {NoStop}%
\bibitem [{\citenamefont {Chen}\ \emph {et~al.}(2015)\citenamefont {Chen},
  \citenamefont {Hu}, \citenamefont {Deng},\ and\ \citenamefont
  {Xing}}]{C4CS00461B}%
  \BibitemOpen
  \bibfield  {author} {\bibinfo {author} {\bibfnamefont {J.}~\bibnamefont
  {Chen}}, \bibinfo {author} {\bibfnamefont {L.}~\bibnamefont {Hu}}, \bibinfo
  {author} {\bibfnamefont {J.}~\bibnamefont {Deng}}, \ and\ \bibinfo {author}
  {\bibfnamefont {X.}~\bibnamefont {Xing}},\ }\href {\doibase
  10.1039/C4CS00461B} {\bibfield  {journal} {\bibinfo  {journal} {Chem. Soc.
  Rev.}\ }\textbf {\bibinfo {volume} {44}},\ \bibinfo {pages} {3522} (\bibinfo
  {year} {2015})}\BibitemShut {NoStop}%
\bibitem [{\citenamefont {Chatterji}\ \emph {et~al.}(2008)\citenamefont
  {Chatterji}, \citenamefont {Henry}, \citenamefont {Mittal},\ and\
  \citenamefont {Chaplot}}]{PhysRevB.78.134105}%
  \BibitemOpen
  \bibfield  {author} {\bibinfo {author} {\bibfnamefont {T.}~\bibnamefont
  {Chatterji}}, \bibinfo {author} {\bibfnamefont {P.~F.}\ \bibnamefont
  {Henry}}, \bibinfo {author} {\bibfnamefont {R.}~\bibnamefont {Mittal}}, \
  and\ \bibinfo {author} {\bibfnamefont {S.~L.}\ \bibnamefont {Chaplot}},\
  }\href {\doibase 10.1103/PhysRevB.78.134105} {\bibfield  {journal} {\bibinfo
  {journal} {Phys. Rev. B}\ }\textbf {\bibinfo {volume} {78}},\ \bibinfo
  {pages} {134105} (\bibinfo {year} {2008})}\BibitemShut {NoStop}%
\bibitem [{\citenamefont {García-Muñoz}\ \emph {et~al.}(1997)\citenamefont
  {García-Muñoz}, \citenamefont {Suaaidi},\ and\ \citenamefont
  {Ritter}}]{GARCIAMUNOZ1997854}%
  \BibitemOpen
  \bibfield  {author} {\bibinfo {author} {\bibfnamefont {J.}~\bibnamefont
  {García-Muñoz}}, \bibinfo {author} {\bibfnamefont {M.}~\bibnamefont
  {Suaaidi}}, \ and\ \bibinfo {author} {\bibfnamefont {C.}~\bibnamefont
  {Ritter}},\ }\href {\doibase https://doi.org/10.1016/S0921-4526(96)01126-X}
  {\bibfield  {journal} {\bibinfo  {journal} {Physica B: Condensed Matter}\
  }\textbf {\bibinfo {volume} {234-236}},\ \bibinfo {pages} {854 } (\bibinfo
  {year} {1997})},\ \bibinfo {note} {proceedings of the First European
  Conference on Neutron Scattering}\BibitemShut {NoStop}%
\bibitem [{\citenamefont {Huang}\ \emph {et~al.}(1997)\citenamefont {Huang},
  \citenamefont {Santoro}, \citenamefont {Lynn}, \citenamefont {Erwin},
  \citenamefont {Borchers}, \citenamefont {Peng},\ and\ \citenamefont
  {Greene}}]{PhysRevB.55.14987}%
  \BibitemOpen
  \bibfield  {author} {\bibinfo {author} {\bibfnamefont {Q.}~\bibnamefont
  {Huang}}, \bibinfo {author} {\bibfnamefont {A.}~\bibnamefont {Santoro}},
  \bibinfo {author} {\bibfnamefont {J.~W.}\ \bibnamefont {Lynn}}, \bibinfo
  {author} {\bibfnamefont {R.~W.}\ \bibnamefont {Erwin}}, \bibinfo {author}
  {\bibfnamefont {J.~A.}\ \bibnamefont {Borchers}}, \bibinfo {author}
  {\bibfnamefont {J.~L.}\ \bibnamefont {Peng}}, \ and\ \bibinfo {author}
  {\bibfnamefont {R.~L.}\ \bibnamefont {Greene}},\ }\href {\doibase
  10.1103/PhysRevB.55.14987} {\bibfield  {journal} {\bibinfo  {journal} {Phys.
  Rev. B}\ }\textbf {\bibinfo {volume} {55}},\ \bibinfo {pages} {14987}
  (\bibinfo {year} {1997})}\BibitemShut {NoStop}%
\bibitem [{\citenamefont {Kiyama}\ \emph {et~al.}(1996)\citenamefont {Kiyama},
  \citenamefont {Yoshimura}, \citenamefont {Kosuge}, \citenamefont {Ikeda},\
  and\ \citenamefont {Bando}}]{PhysRevB.54.R756}%
  \BibitemOpen
  \bibfield  {author} {\bibinfo {author} {\bibfnamefont {T.}~\bibnamefont
  {Kiyama}}, \bibinfo {author} {\bibfnamefont {K.}~\bibnamefont {Yoshimura}},
  \bibinfo {author} {\bibfnamefont {K.}~\bibnamefont {Kosuge}}, \bibinfo
  {author} {\bibfnamefont {Y.}~\bibnamefont {Ikeda}}, \ and\ \bibinfo {author}
  {\bibfnamefont {Y.}~\bibnamefont {Bando}},\ }\href {\doibase
  10.1103/PhysRevB.54.R756} {\bibfield  {journal} {\bibinfo  {journal} {Phys.
  Rev. B}\ }\textbf {\bibinfo {volume} {54}},\ \bibinfo {pages} {R756}
  (\bibinfo {year} {1996})}\BibitemShut {NoStop}%
\bibitem [{\citenamefont {Dove}\ \emph {et~al.}(1995)\citenamefont {Dove},
  \citenamefont {Heine},\ and\ \citenamefont
  {Hammonds}}]{dove_heine_hammonds_1995}%
  \BibitemOpen
  \bibfield  {author} {\bibinfo {author} {\bibfnamefont {M.~T.}\ \bibnamefont
  {Dove}}, \bibinfo {author} {\bibfnamefont {V.}~\bibnamefont {Heine}}, \ and\
  \bibinfo {author} {\bibfnamefont {K.~D.}\ \bibnamefont {Hammonds}},\ }\href
  {\doibase 10.1180/minmag.1995.059.397.07} {\bibfield  {journal} {\bibinfo
  {journal} {Mineralogical Magazine}\ }\textbf {\bibinfo {volume} {59}},\
  \bibinfo {pages} {629–639} (\bibinfo {year} {1995})}\BibitemShut {NoStop}%
\bibitem [{\citenamefont {Takenaka}\ \emph {et~al.}(2014)\citenamefont
  {Takenaka}, \citenamefont {Ichigo}, \citenamefont {Hamada}, \citenamefont
  {Ozawa}, \citenamefont {Shibayama}, \citenamefont {Inagaki},\ and\
  \citenamefont {Asano}}]{doi:10.1088/1468-6996/15/1/015009}%
  \BibitemOpen
  \bibfield  {author} {\bibinfo {author} {\bibfnamefont {K.}~\bibnamefont
  {Takenaka}}, \bibinfo {author} {\bibfnamefont {M.}~\bibnamefont {Ichigo}},
  \bibinfo {author} {\bibfnamefont {T.}~\bibnamefont {Hamada}}, \bibinfo
  {author} {\bibfnamefont {A.}~\bibnamefont {Ozawa}}, \bibinfo {author}
  {\bibfnamefont {T.}~\bibnamefont {Shibayama}}, \bibinfo {author}
  {\bibfnamefont {T.}~\bibnamefont {Inagaki}}, \ and\ \bibinfo {author}
  {\bibfnamefont {K.}~\bibnamefont {Asano}},\ }\href {\doibase
  10.1088/1468-6996/15/1/015009} {\bibfield  {journal} {\bibinfo  {journal}
  {Science and Technology of Advanced Materials}\ }\textbf {\bibinfo {volume}
  {15}},\ \bibinfo {pages} {015009} (\bibinfo {year} {2014})}\BibitemShut
  {NoStop}%
\bibitem [{\citenamefont {Saha}\ \emph {et~al.}(2011)\citenamefont {Saha},
  \citenamefont {Shireen}, \citenamefont {Bera}, \citenamefont {Shirodkar},
  \citenamefont {Sundarayya}, \citenamefont {Kalarikkal}, \citenamefont
  {Yusuf}, \citenamefont {Waghmare}, \citenamefont {Sundaresan},\ and\
  \citenamefont {Rao}}]{SAHA2011494}%
  \BibitemOpen
  \bibfield  {author} {\bibinfo {author} {\bibfnamefont {R.}~\bibnamefont
  {Saha}}, \bibinfo {author} {\bibfnamefont {A.}~\bibnamefont {Shireen}},
  \bibinfo {author} {\bibfnamefont {A.}~\bibnamefont {Bera}}, \bibinfo {author}
  {\bibfnamefont {S.~N.}\ \bibnamefont {Shirodkar}}, \bibinfo {author}
  {\bibfnamefont {Y.}~\bibnamefont {Sundarayya}}, \bibinfo {author}
  {\bibfnamefont {N.}~\bibnamefont {Kalarikkal}}, \bibinfo {author}
  {\bibfnamefont {S.}~\bibnamefont {Yusuf}}, \bibinfo {author} {\bibfnamefont
  {U.~V.}\ \bibnamefont {Waghmare}}, \bibinfo {author} {\bibfnamefont
  {A.}~\bibnamefont {Sundaresan}}, \ and\ \bibinfo {author} {\bibfnamefont
  {C.}~\bibnamefont {Rao}},\ }\href {\doibase
  https://doi.org/10.1016/j.jssc.2010.12.025} {\bibfield  {journal} {\bibinfo
  {journal} {Journal of Solid State Chemistry}\ }\textbf {\bibinfo {volume}
  {184}},\ \bibinfo {pages} {494 } (\bibinfo {year} {2011})}\BibitemShut
  {NoStop}%
\bibitem [{\citenamefont {Takenaka}\ \emph {et~al.}(2017)\citenamefont
  {Takenaka}, \citenamefont {Okamoto}, \citenamefont {Shinoda}, \citenamefont
  {Katayama},\ and\ \citenamefont {Sakai}}]{takenaka2017colossal}%
  \BibitemOpen
  \bibfield  {author} {\bibinfo {author} {\bibfnamefont {K.}~\bibnamefont
  {Takenaka}}, \bibinfo {author} {\bibfnamefont {Y.}~\bibnamefont {Okamoto}},
  \bibinfo {author} {\bibfnamefont {T.}~\bibnamefont {Shinoda}}, \bibinfo
  {author} {\bibfnamefont {N.}~\bibnamefont {Katayama}}, \ and\ \bibinfo
  {author} {\bibfnamefont {Y.}~\bibnamefont {Sakai}},\ }\href {\doibase
  10.1038/ncomms14102} {\bibfield  {journal} {\bibinfo  {journal} {Nature
  communications}\ }\textbf {\bibinfo {volume} {8}},\ \bibinfo {pages} {1}
  (\bibinfo {year} {2017})}\BibitemShut {NoStop}%
\bibitem [{\citenamefont {Mary}\ \emph {et~al.}(1996)\citenamefont {Mary},
  \citenamefont {Evans}, \citenamefont {Vogt},\ and\ \citenamefont
  {Sleight}}]{Mary90}%
  \BibitemOpen
  \bibfield  {author} {\bibinfo {author} {\bibfnamefont {T.~A.}\ \bibnamefont
  {Mary}}, \bibinfo {author} {\bibfnamefont {J.~S.~O.}\ \bibnamefont {Evans}},
  \bibinfo {author} {\bibfnamefont {T.}~\bibnamefont {Vogt}}, \ and\ \bibinfo
  {author} {\bibfnamefont {A.~W.}\ \bibnamefont {Sleight}},\ }\href {\doibase
  10.1126/science.272.5258.90} {\bibfield  {journal} {\bibinfo  {journal}
  {Science}\ }\textbf {\bibinfo {volume} {272}},\ \bibinfo {pages} {90}
  (\bibinfo {year} {1996})}\BibitemShut {NoStop}%
\bibitem [{\citenamefont {Ernst}\ \emph {et~al.}(1998)\citenamefont {Ernst},
  \citenamefont {Broholm}, \citenamefont {Kowach},\ and\ \citenamefont
  {Ramirez}}]{ernst1998phonon}%
  \BibitemOpen
  \bibfield  {author} {\bibinfo {author} {\bibfnamefont {G.}~\bibnamefont
  {Ernst}}, \bibinfo {author} {\bibfnamefont {C.}~\bibnamefont {Broholm}},
  \bibinfo {author} {\bibfnamefont {G.}~\bibnamefont {Kowach}}, \ and\ \bibinfo
  {author} {\bibfnamefont {A.}~\bibnamefont {Ramirez}},\ }\href {\doibase
  10.1038/24115} {\bibfield  {journal} {\bibinfo  {journal} {Nature}\ }\textbf
  {\bibinfo {volume} {396}},\ \bibinfo {pages} {147} (\bibinfo {year}
  {1998})}\BibitemShut {NoStop}%
\bibitem [{\citenamefont {Hemberger}\ \emph {et~al.}(2007)\citenamefont
  {Hemberger}, \citenamefont {von Nidda}, \citenamefont {Tsurkan},\ and\
  \citenamefont {Loidl}}]{PhysRevLett.98.147203}%
  \BibitemOpen
  \bibfield  {author} {\bibinfo {author} {\bibfnamefont {J.}~\bibnamefont
  {Hemberger}}, \bibinfo {author} {\bibfnamefont {H.-A.~K.}\ \bibnamefont {von
  Nidda}}, \bibinfo {author} {\bibfnamefont {V.}~\bibnamefont {Tsurkan}}, \
  and\ \bibinfo {author} {\bibfnamefont {A.}~\bibnamefont {Loidl}},\ }\href
  {\doibase 10.1103/PhysRevLett.98.147203} {\bibfield  {journal} {\bibinfo
  {journal} {Phys. Rev. Lett.}\ }\textbf {\bibinfo {volume} {98}},\ \bibinfo
  {pages} {147203} (\bibinfo {year} {2007})}\BibitemShut {NoStop}%
\bibitem [{\citenamefont {Pokharel}\ \emph {et~al.}(2018)\citenamefont
  {Pokharel}, \citenamefont {May}, \citenamefont {Parker}, \citenamefont
  {Calder}, \citenamefont {Ehlers}, \citenamefont {Huq}, \citenamefont
  {Kimber}, \citenamefont {Arachchige}, \citenamefont {Poudel}, \citenamefont
  {McGuire}, \citenamefont {Mandrus},\ and\ \citenamefont
  {Christianson}}]{PhysRevB.97.134117}%
  \BibitemOpen
  \bibfield  {author} {\bibinfo {author} {\bibfnamefont {G.}~\bibnamefont
  {Pokharel}}, \bibinfo {author} {\bibfnamefont {A.~F.}\ \bibnamefont {May}},
  \bibinfo {author} {\bibfnamefont {D.~S.}\ \bibnamefont {Parker}}, \bibinfo
  {author} {\bibfnamefont {S.}~\bibnamefont {Calder}}, \bibinfo {author}
  {\bibfnamefont {G.}~\bibnamefont {Ehlers}}, \bibinfo {author} {\bibfnamefont
  {A.}~\bibnamefont {Huq}}, \bibinfo {author} {\bibfnamefont {S.~A.~J.}\
  \bibnamefont {Kimber}}, \bibinfo {author} {\bibfnamefont {H.~S.}\
  \bibnamefont {Arachchige}}, \bibinfo {author} {\bibfnamefont
  {L.}~\bibnamefont {Poudel}}, \bibinfo {author} {\bibfnamefont {M.~A.}\
  \bibnamefont {McGuire}}, \bibinfo {author} {\bibfnamefont {D.}~\bibnamefont
  {Mandrus}}, \ and\ \bibinfo {author} {\bibfnamefont {A.~D.}\ \bibnamefont
  {Christianson}},\ }\href {\doibase 10.1103/PhysRevB.97.134117} {\bibfield
  {journal} {\bibinfo  {journal} {Phys. Rev. B}\ }\textbf {\bibinfo {volume}
  {97}},\ \bibinfo {pages} {134117} (\bibinfo {year} {2018})}\BibitemShut
  {NoStop}%
\bibitem [{\citenamefont {Villaescusa}\ \emph {et~al.}(2001)\citenamefont
  {Villaescusa}, \citenamefont {Lightfoot}, \citenamefont {Teat},\ and\
  \citenamefont {Morris}}]{villaescusa2001variable}%
  \BibitemOpen
  \bibfield  {author} {\bibinfo {author} {\bibfnamefont {L.~A.}\ \bibnamefont
  {Villaescusa}}, \bibinfo {author} {\bibfnamefont {P.}~\bibnamefont
  {Lightfoot}}, \bibinfo {author} {\bibfnamefont {S.~J.}\ \bibnamefont {Teat}},
  \ and\ \bibinfo {author} {\bibfnamefont {R.~E.}\ \bibnamefont {Morris}},\
  }\href {\doibase 10.1021/ja015797o} {\bibfield  {journal} {\bibinfo
  {journal} {Journal of the American Chemical Society}\ }\textbf {\bibinfo
  {volume} {123}},\ \bibinfo {pages} {5453} (\bibinfo {year}
  {2001})}\BibitemShut {NoStop}%
\bibitem [{\citenamefont {Woodcock}\ \emph {et~al.}(1999)\citenamefont
  {Woodcock}, \citenamefont {Lightfoot}, \citenamefont {Villaescusa},
  \citenamefont {D{\'\i}az-Caba{\~n}as}, \citenamefont {Camblor},\ and\
  \citenamefont {Engberg}}]{woodcock1999negative}%
  \BibitemOpen
  \bibfield  {author} {\bibinfo {author} {\bibfnamefont {D.~A.}\ \bibnamefont
  {Woodcock}}, \bibinfo {author} {\bibfnamefont {P.}~\bibnamefont {Lightfoot}},
  \bibinfo {author} {\bibfnamefont {L.~A.}\ \bibnamefont {Villaescusa}},
  \bibinfo {author} {\bibfnamefont {M.-J.}\ \bibnamefont
  {D{\'\i}az-Caba{\~n}as}}, \bibinfo {author} {\bibfnamefont {M.~A.}\
  \bibnamefont {Camblor}}, \ and\ \bibinfo {author} {\bibfnamefont
  {D.}~\bibnamefont {Engberg}},\ }\href {\doibase 10.1021/cm991047q} {\bibfield
   {journal} {\bibinfo  {journal} {Chemistry of materials}\ }\textbf {\bibinfo
  {volume} {11}},\ \bibinfo {pages} {2508} (\bibinfo {year}
  {1999})}\BibitemShut {NoStop}%
\bibitem [{\citenamefont {Rau}\ \emph {et~al.}(2016)\citenamefont {Rau},
  \citenamefont {Lee},\ and\ \citenamefont
  {Kee}}]{doi:10.1146/annurev-conmatphys-031115-011319}%
  \BibitemOpen
  \bibfield  {author} {\bibinfo {author} {\bibfnamefont {J.~G.}\ \bibnamefont
  {Rau}}, \bibinfo {author} {\bibfnamefont {E.~K.-H.}\ \bibnamefont {Lee}}, \
  and\ \bibinfo {author} {\bibfnamefont {H.-Y.}\ \bibnamefont {Kee}},\ }\href
  {\doibase 10.1146/annurev-conmatphys-031115-011319} {\bibfield  {journal}
  {\bibinfo  {journal} {Annual Review of Condensed Matter Physics}\ }\textbf
  {\bibinfo {volume} {7}},\ \bibinfo {pages} {195} (\bibinfo {year}
  {2016})}\BibitemShut {NoStop}%
\bibitem [{\citenamefont {Zhou}\ \emph {et~al.}(2012)\citenamefont {Zhou},
  \citenamefont {Kiswandhi}, \citenamefont {Barlas}, \citenamefont {Brooks},
  \citenamefont {Siegrist}, \citenamefont {Li}, \citenamefont {Balicas},
  \citenamefont {Cheng},\ and\ \citenamefont {Rivadulla}}]{PhysRevB.85.041201}%
  \BibitemOpen
  \bibfield  {author} {\bibinfo {author} {\bibfnamefont {H.~D.}\ \bibnamefont
  {Zhou}}, \bibinfo {author} {\bibfnamefont {A.}~\bibnamefont {Kiswandhi}},
  \bibinfo {author} {\bibfnamefont {Y.}~\bibnamefont {Barlas}}, \bibinfo
  {author} {\bibfnamefont {J.~S.}\ \bibnamefont {Brooks}}, \bibinfo {author}
  {\bibfnamefont {T.}~\bibnamefont {Siegrist}}, \bibinfo {author}
  {\bibfnamefont {G.}~\bibnamefont {Li}}, \bibinfo {author} {\bibfnamefont
  {L.}~\bibnamefont {Balicas}}, \bibinfo {author} {\bibfnamefont {J.~G.}\
  \bibnamefont {Cheng}}, \ and\ \bibinfo {author} {\bibfnamefont
  {F.}~\bibnamefont {Rivadulla}},\ }\href {\doibase 10.1103/PhysRevB.85.041201}
  {\bibfield  {journal} {\bibinfo  {journal} {Phys. Rev. B}\ }\textbf {\bibinfo
  {volume} {85}},\ \bibinfo {pages} {041201} (\bibinfo {year}
  {2012})}\BibitemShut {NoStop}%
\bibitem [{\citenamefont {Kimber}\ \emph {et~al.}(2012)\citenamefont {Kimber},
  \citenamefont {Senn}, \citenamefont {Fratini}, \citenamefont {Wu},
  \citenamefont {Hill}, \citenamefont {Manuel}, \citenamefont {Attfield},
  \citenamefont {Argyriou},\ and\ \citenamefont
  {Henry}}]{PhysRevLett.108.217205}%
  \BibitemOpen
  \bibfield  {author} {\bibinfo {author} {\bibfnamefont {S.~A.~J.}\
  \bibnamefont {Kimber}}, \bibinfo {author} {\bibfnamefont {M.~S.}\
  \bibnamefont {Senn}}, \bibinfo {author} {\bibfnamefont {S.}~\bibnamefont
  {Fratini}}, \bibinfo {author} {\bibfnamefont {H.}~\bibnamefont {Wu}},
  \bibinfo {author} {\bibfnamefont {A.~H.}\ \bibnamefont {Hill}}, \bibinfo
  {author} {\bibfnamefont {P.}~\bibnamefont {Manuel}}, \bibinfo {author}
  {\bibfnamefont {J.~P.}\ \bibnamefont {Attfield}}, \bibinfo {author}
  {\bibfnamefont {D.~N.}\ \bibnamefont {Argyriou}}, \ and\ \bibinfo {author}
  {\bibfnamefont {P.~F.}\ \bibnamefont {Henry}},\ }\href {\doibase
  10.1103/PhysRevLett.108.217205} {\bibfield  {journal} {\bibinfo  {journal}
  {Phys. Rev. Lett.}\ }\textbf {\bibinfo {volume} {108}},\ \bibinfo {pages}
  {217205} (\bibinfo {year} {2012})}\BibitemShut {NoStop}%
\bibitem [{\citenamefont {Yokota}\ \emph {et~al.}(2014)\citenamefont {Yokota},
  \citenamefont {Kurita},\ and\ \citenamefont {Tanaka}}]{PhysRevB.90.014403}%
  \BibitemOpen
  \bibfield  {author} {\bibinfo {author} {\bibfnamefont {K.}~\bibnamefont
  {Yokota}}, \bibinfo {author} {\bibfnamefont {N.}~\bibnamefont {Kurita}}, \
  and\ \bibinfo {author} {\bibfnamefont {H.}~\bibnamefont {Tanaka}},\ }\href
  {\doibase 10.1103/PhysRevB.90.014403} {\bibfield  {journal} {\bibinfo
  {journal} {Phys. Rev. B}\ }\textbf {\bibinfo {volume} {90}},\ \bibinfo
  {pages} {014403} (\bibinfo {year} {2014})}\BibitemShut {NoStop}%
\bibitem [{\citenamefont {Lee}\ \emph {et~al.}(2016)\citenamefont {Lee},
  \citenamefont {Choi}, \citenamefont {Ma}, \citenamefont {Sinclair},
  \citenamefont {Cruz},\ and\ \citenamefont {Zhou}}]{Lee_2016}%
  \BibitemOpen
  \bibfield  {author} {\bibinfo {author} {\bibfnamefont {M.}~\bibnamefont
  {Lee}}, \bibinfo {author} {\bibfnamefont {E.~S.}\ \bibnamefont {Choi}},
  \bibinfo {author} {\bibfnamefont {J.}~\bibnamefont {Ma}}, \bibinfo {author}
  {\bibfnamefont {R.}~\bibnamefont {Sinclair}}, \bibinfo {author}
  {\bibfnamefont {C.~R.~D.}\ \bibnamefont {Cruz}}, \ and\ \bibinfo {author}
  {\bibfnamefont {H.~D.}\ \bibnamefont {Zhou}},\ }\href {\doibase
  10.1088/0953-8984/28/47/476004} {\bibfield  {journal} {\bibinfo  {journal}
  {Journal of Physics: Condensed Matter}\ }\textbf {\bibinfo {volume} {28}},\
  \bibinfo {pages} {476004} (\bibinfo {year} {2016})}\BibitemShut {NoStop}%
\bibitem [{\citenamefont {Ranjith}\ \emph {et~al.}(2017)\citenamefont
  {Ranjith}, \citenamefont {Brinda}, \citenamefont {Arjun}, \citenamefont
  {Hegde},\ and\ \citenamefont {Nath}}]{Ranjith_2017}%
  \BibitemOpen
  \bibfield  {author} {\bibinfo {author} {\bibfnamefont {K.~M.}\ \bibnamefont
  {Ranjith}}, \bibinfo {author} {\bibfnamefont {K.}~\bibnamefont {Brinda}},
  \bibinfo {author} {\bibfnamefont {U.}~\bibnamefont {Arjun}}, \bibinfo
  {author} {\bibfnamefont {N.~G.}\ \bibnamefont {Hegde}}, \ and\ \bibinfo
  {author} {\bibfnamefont {R.}~\bibnamefont {Nath}},\ }\href {\doibase
  10.1088/1361-648x/aa57be} {\bibfield  {journal} {\bibinfo  {journal} {Journal
  of Physics: Condensed Matter}\ }\textbf {\bibinfo {volume} {29}},\ \bibinfo
  {pages} {115804} (\bibinfo {year} {2017})}\BibitemShut {NoStop}%
\bibitem [{\citenamefont {Garg}\ \emph {et~al.}(2020)\citenamefont {Garg},
  \citenamefont {Roy}, \citenamefont {Lonsky}, \citenamefont {Manuel},
  \citenamefont {Cervellino}, \citenamefont {Müller}, \citenamefont {Kabir},\
  and\ \citenamefont {Nair}}]{garg2020evolution}%
  \BibitemOpen
  \bibfield  {author} {\bibinfo {author} {\bibfnamefont {C.}~\bibnamefont
  {Garg}}, \bibinfo {author} {\bibfnamefont {D.}~\bibnamefont {Roy}}, \bibinfo
  {author} {\bibfnamefont {M.}~\bibnamefont {Lonsky}}, \bibinfo {author}
  {\bibfnamefont {P.}~\bibnamefont {Manuel}}, \bibinfo {author} {\bibfnamefont
  {A.}~\bibnamefont {Cervellino}}, \bibinfo {author} {\bibfnamefont
  {J.}~\bibnamefont {Müller}}, \bibinfo {author} {\bibfnamefont
  {M.}~\bibnamefont {Kabir}}, \ and\ \bibinfo {author} {\bibfnamefont
  {S.}~\bibnamefont {Nair}},\ }\href@noop {} {} (\bibinfo {year} {2020}),\
  \Eprint {http://arxiv.org/abs/2009.13822} {arXiv:2009.13822
  [cond-mat.str-el]} \BibitemShut {NoStop}%
\bibitem [{\citenamefont {Willmott}\ \emph {et~al.}(2013)\citenamefont
  {Willmott}, \citenamefont {Meister}, \citenamefont {Leake}, \citenamefont
  {Lange}, \citenamefont {Bergamaschi}, \citenamefont {B{\"{o}}ge},
  \citenamefont {Calvi}, \citenamefont {Cancellieri}, \citenamefont {Casati},
  \citenamefont {Cervellino}, \citenamefont {Chen}, \citenamefont {David},
  \citenamefont {Flechsig}, \citenamefont {Gozzo}, \citenamefont {Henrich},
  \citenamefont {J{\"{a}}ggi-Spielmann}, \citenamefont {Jakob}, \citenamefont
  {Kalichava}, \citenamefont {Karvinen}, \citenamefont {Krempasky},
  \citenamefont {L{\"{u}}deke}, \citenamefont {L{\"{u}}scher}, \citenamefont
  {Maag}, \citenamefont {Quitmann}, \citenamefont {Reinle-Schmitt},
  \citenamefont {Schmidt}, \citenamefont {Schmitt}, \citenamefont {Streun},
  \citenamefont {Vartiainen}, \citenamefont {Vitins}, \citenamefont {Wang},\
  and\ \citenamefont {Wullschleger}}]{SLS}%
  \BibitemOpen
  \bibfield  {author} {\bibinfo {author} {\bibfnamefont {P.~R.}\ \bibnamefont
  {Willmott}}, \bibinfo {author} {\bibfnamefont {D.}~\bibnamefont {Meister}},
  \bibinfo {author} {\bibfnamefont {S.~J.}\ \bibnamefont {Leake}}, \bibinfo
  {author} {\bibfnamefont {M.}~\bibnamefont {Lange}}, \bibinfo {author}
  {\bibfnamefont {A.}~\bibnamefont {Bergamaschi}}, \bibinfo {author}
  {\bibfnamefont {M.}~\bibnamefont {B{\"{o}}ge}}, \bibinfo {author}
  {\bibfnamefont {M.}~\bibnamefont {Calvi}}, \bibinfo {author} {\bibfnamefont
  {C.}~\bibnamefont {Cancellieri}}, \bibinfo {author} {\bibfnamefont
  {N.}~\bibnamefont {Casati}}, \bibinfo {author} {\bibfnamefont
  {A.}~\bibnamefont {Cervellino}}, \bibinfo {author} {\bibfnamefont
  {Q.}~\bibnamefont {Chen}}, \bibinfo {author} {\bibfnamefont {C.}~\bibnamefont
  {David}}, \bibinfo {author} {\bibfnamefont {U.}~\bibnamefont {Flechsig}},
  \bibinfo {author} {\bibfnamefont {F.}~\bibnamefont {Gozzo}}, \bibinfo
  {author} {\bibfnamefont {B.}~\bibnamefont {Henrich}}, \bibinfo {author}
  {\bibfnamefont {S.}~\bibnamefont {J{\"{a}}ggi-Spielmann}}, \bibinfo {author}
  {\bibfnamefont {B.}~\bibnamefont {Jakob}}, \bibinfo {author} {\bibfnamefont
  {I.}~\bibnamefont {Kalichava}}, \bibinfo {author} {\bibfnamefont
  {P.}~\bibnamefont {Karvinen}}, \bibinfo {author} {\bibfnamefont
  {J.}~\bibnamefont {Krempasky}}, \bibinfo {author} {\bibfnamefont
  {A.}~\bibnamefont {L{\"{u}}deke}}, \bibinfo {author} {\bibfnamefont
  {R.}~\bibnamefont {L{\"{u}}scher}}, \bibinfo {author} {\bibfnamefont
  {S.}~\bibnamefont {Maag}}, \bibinfo {author} {\bibfnamefont {C.}~\bibnamefont
  {Quitmann}}, \bibinfo {author} {\bibfnamefont {M.~L.}\ \bibnamefont
  {Reinle-Schmitt}}, \bibinfo {author} {\bibfnamefont {T.}~\bibnamefont
  {Schmidt}}, \bibinfo {author} {\bibfnamefont {B.}~\bibnamefont {Schmitt}},
  \bibinfo {author} {\bibfnamefont {A.}~\bibnamefont {Streun}}, \bibinfo
  {author} {\bibfnamefont {I.}~\bibnamefont {Vartiainen}}, \bibinfo {author}
  {\bibfnamefont {M.}~\bibnamefont {Vitins}}, \bibinfo {author} {\bibfnamefont
  {X.}~\bibnamefont {Wang}}, \ and\ \bibinfo {author} {\bibfnamefont
  {R.}~\bibnamefont {Wullschleger}},\ }\href {\doibase
  10.1107/S0909049513018475} {\bibfield  {journal} {\bibinfo  {journal}
  {Journal of Synchrotron Radiation}\ }\textbf {\bibinfo {volume} {20}},\
  \bibinfo {pages} {667} (\bibinfo {year} {2013})}\BibitemShut {NoStop}%
\bibitem [{\citenamefont {Rodriguez-Carvajal}(2001)}]{Fullprof}%
  \BibitemOpen
  \bibfield  {author} {\bibinfo {author} {\bibfnamefont {J.}~\bibnamefont
  {Rodriguez-Carvajal}},\ }\href@noop {} {\emph {\bibinfo {title} {An
  introduction to the programme FULLPROF}}}\ (\bibinfo  {publisher}
  {Laboratoire Leon Brillouin , CEA-CNRS ,Saclay, France},\ \bibinfo {year}
  {2001})\BibitemShut {NoStop}%
\bibitem [{\citenamefont {Momma}\ and\ \citenamefont {Izumi}(2011)}]{vesta}%
  \BibitemOpen
  \bibfield  {author} {\bibinfo {author} {\bibfnamefont {K.}~\bibnamefont
  {Momma}}\ and\ \bibinfo {author} {\bibfnamefont {F.}~\bibnamefont {Izumi}},\
  }\href {\doibase 10.1107/S0021889811038970} {\bibfield  {journal} {\bibinfo
  {journal} {J. Appl. Cryst.}\ }\textbf {\bibinfo {volume} {44}},\ \bibinfo
  {pages} {1272} (\bibinfo {year} {2011})}\BibitemShut {NoStop}%
\bibitem [{\citenamefont {Treiber}\ \emph {et~al.}(1982)\citenamefont
  {Treiber}, \citenamefont {Kemmler-Sack},\ and\ \citenamefont
  {Ehmann}}]{hexa}%
  \BibitemOpen
  \bibfield  {author} {\bibinfo {author} {\bibfnamefont {U.}~\bibnamefont
  {Treiber}}, \bibinfo {author} {\bibfnamefont {S.}~\bibnamefont
  {Kemmler-Sack}}, \ and\ \bibinfo {author} {\bibfnamefont {A.}~\bibnamefont
  {Ehmann}},\ }\href {\doibase 10.1002/zaac.19824870117} {\bibfield  {journal}
  {\bibinfo  {journal} {Zeitschrift für anorganische und allgemeine Chemie}\
  }\textbf {\bibinfo {volume} {487}},\ \bibinfo {pages} {189} (\bibinfo {year}
  {1982})}\BibitemShut {NoStop}%
\bibitem [{\citenamefont {{Sakamoto}}\ \emph {et~al.}(2006)\citenamefont
  {{Sakamoto}}, \citenamefont {{Doi}},\ and\ \citenamefont {{Hinatsu}}}]{IrIr}%
  \BibitemOpen
  \bibfield  {author} {\bibinfo {author} {\bibfnamefont {T.}~\bibnamefont
  {{Sakamoto}}}, \bibinfo {author} {\bibfnamefont {Y.}~\bibnamefont {{Doi}}}, \
  and\ \bibinfo {author} {\bibfnamefont {Y.}~\bibnamefont {{Hinatsu}}},\ }\href
  {\doibase 10.1016/j.jssc.2006.04.055} {\bibfield  {journal} {\bibinfo
  {journal} {Journal of Solid State Chemistry France}\ }\textbf {\bibinfo
  {volume} {179}},\ \bibinfo {pages} {2595} (\bibinfo {year}
  {2006})}\BibitemShut {NoStop}%
\bibitem [{\citenamefont {Takenaka}\ and\ \citenamefont
  {Takagi}(2005)}]{doi:10.1063/1.2147726}%
  \BibitemOpen
  \bibfield  {author} {\bibinfo {author} {\bibfnamefont {K.}~\bibnamefont
  {Takenaka}}\ and\ \bibinfo {author} {\bibfnamefont {H.}~\bibnamefont
  {Takagi}},\ }\href {\doibase 10.1063/1.2147726} {\bibfield  {journal}
  {\bibinfo  {journal} {Applied Physics Letters}\ }\textbf {\bibinfo {volume}
  {87}},\ \bibinfo {pages} {261902} (\bibinfo {year} {2005})}\BibitemShut
  {NoStop}%
\bibitem [{\citenamefont {Yamada}\ \emph {et~al.}(2011)\citenamefont {Yamada},
  \citenamefont {Tsuchida}, \citenamefont {Ohgushi}, \citenamefont {Hayashi},
  \citenamefont {Kim}, \citenamefont {Tsuji}, \citenamefont {Takahashi},
  \citenamefont {Matsushita}, \citenamefont {Nishiyama}, \citenamefont {Inoue},
  \citenamefont {Irifune}, \citenamefont {Kato}, \citenamefont {Takata},\ and\
  \citenamefont {Takano}}]{doi:10.1002/anie.201102228}%
  \BibitemOpen
  \bibfield  {author} {\bibinfo {author} {\bibfnamefont {I.}~\bibnamefont
  {Yamada}}, \bibinfo {author} {\bibfnamefont {K.}~\bibnamefont {Tsuchida}},
  \bibinfo {author} {\bibfnamefont {K.}~\bibnamefont {Ohgushi}}, \bibinfo
  {author} {\bibfnamefont {N.}~\bibnamefont {Hayashi}}, \bibinfo {author}
  {\bibfnamefont {J.}~\bibnamefont {Kim}}, \bibinfo {author} {\bibfnamefont
  {N.}~\bibnamefont {Tsuji}}, \bibinfo {author} {\bibfnamefont
  {R.}~\bibnamefont {Takahashi}}, \bibinfo {author} {\bibfnamefont
  {M.}~\bibnamefont {Matsushita}}, \bibinfo {author} {\bibfnamefont
  {N.}~\bibnamefont {Nishiyama}}, \bibinfo {author} {\bibfnamefont
  {T.}~\bibnamefont {Inoue}}, \bibinfo {author} {\bibfnamefont
  {T.}~\bibnamefont {Irifune}}, \bibinfo {author} {\bibfnamefont
  {K.}~\bibnamefont {Kato}}, \bibinfo {author} {\bibfnamefont {M.}~\bibnamefont
  {Takata}}, \ and\ \bibinfo {author} {\bibfnamefont {M.}~\bibnamefont
  {Takano}},\ }\href {\doibase 10.1002/anie.201102228} {\bibfield  {journal}
  {\bibinfo  {journal} {Angewandte Chemie International Edition}\ }\textbf
  {\bibinfo {volume} {50}},\ \bibinfo {pages} {6579} (\bibinfo {year}
  {2011})}\BibitemShut {NoStop}%
\bibitem [{\citenamefont {Filippetti}\ and\ \citenamefont
  {Hill}(2000)}]{PhysRevLett.85.5166}%
  \BibitemOpen
  \bibfield  {author} {\bibinfo {author} {\bibfnamefont {A.}~\bibnamefont
  {Filippetti}}\ and\ \bibinfo {author} {\bibfnamefont {N.~A.}\ \bibnamefont
  {Hill}},\ }\href {\doibase 10.1103/PhysRevLett.85.5166} {\bibfield  {journal}
  {\bibinfo  {journal} {Phys. Rev. Lett.}\ }\textbf {\bibinfo {volume} {85}},\
  \bibinfo {pages} {5166} (\bibinfo {year} {2000})}\BibitemShut {NoStop}%
\bibitem [{\citenamefont {Azuma}\ \emph {et~al.}(2011)\citenamefont {Azuma},
  \citenamefont {Chen}, \citenamefont {Seki}, \citenamefont {Czapski},
  \citenamefont {Olga}, \citenamefont {Oka}, \citenamefont {Mizumaki},
  \citenamefont {Watanuki}, \citenamefont {Ishimatsu}, \citenamefont {Kawamura}
  \emph {et~al.}}]{azuma2011colossal}%
  \BibitemOpen
  \bibfield  {author} {\bibinfo {author} {\bibfnamefont {M.}~\bibnamefont
  {Azuma}}, \bibinfo {author} {\bibfnamefont {W.-t.}\ \bibnamefont {Chen}},
  \bibinfo {author} {\bibfnamefont {H.}~\bibnamefont {Seki}}, \bibinfo {author}
  {\bibfnamefont {M.}~\bibnamefont {Czapski}}, \bibinfo {author} {\bibfnamefont
  {S.}~\bibnamefont {Olga}}, \bibinfo {author} {\bibfnamefont {K.}~\bibnamefont
  {Oka}}, \bibinfo {author} {\bibfnamefont {M.}~\bibnamefont {Mizumaki}},
  \bibinfo {author} {\bibfnamefont {T.}~\bibnamefont {Watanuki}}, \bibinfo
  {author} {\bibfnamefont {N.}~\bibnamefont {Ishimatsu}}, \bibinfo {author}
  {\bibfnamefont {N.}~\bibnamefont {Kawamura}},  \emph {et~al.},\ }\href
  {\doibase 10.1038/ncomms1361} {\bibfield  {journal} {\bibinfo  {journal}
  {Nature communications}\ }\textbf {\bibinfo {volume} {2}},\ \bibinfo {pages}
  {1} (\bibinfo {year} {2011})}\BibitemShut {NoStop}%
\bibitem [{\citenamefont {Chen}\ \emph {et~al.}(2013)\citenamefont {Chen},
  \citenamefont {Wang}, \citenamefont {Huang}, \citenamefont {Hu},
  \citenamefont {Song}, \citenamefont {Deng}, \citenamefont {Yu},\ and\
  \citenamefont {Xing}}]{chen2013effectively}%
  \BibitemOpen
  \bibfield  {author} {\bibinfo {author} {\bibfnamefont {J.}~\bibnamefont
  {Chen}}, \bibinfo {author} {\bibfnamefont {F.}~\bibnamefont {Wang}}, \bibinfo
  {author} {\bibfnamefont {Q.}~\bibnamefont {Huang}}, \bibinfo {author}
  {\bibfnamefont {L.}~\bibnamefont {Hu}}, \bibinfo {author} {\bibfnamefont
  {X.}~\bibnamefont {Song}}, \bibinfo {author} {\bibfnamefont {J.}~\bibnamefont
  {Deng}}, \bibinfo {author} {\bibfnamefont {R.}~\bibnamefont {Yu}}, \ and\
  \bibinfo {author} {\bibfnamefont {X.}~\bibnamefont {Xing}},\ }\href {\doibase
  10.1038/srep02458} {\bibfield  {journal} {\bibinfo  {journal} {Scientific
  reports}\ }\textbf {\bibinfo {volume} {3}},\ \bibinfo {pages} {2458}
  (\bibinfo {year} {2013})}\BibitemShut {NoStop}%
\bibitem [{\citenamefont {Qi}\ \emph {et~al.}(2012)\citenamefont {Qi},
  \citenamefont {Korneta}, \citenamefont {Parkin}, \citenamefont {Hu},\ and\
  \citenamefont {Cao}}]{PhysRevB.85.165143}%
  \BibitemOpen
  \bibfield  {author} {\bibinfo {author} {\bibfnamefont {T.~F.}\ \bibnamefont
  {Qi}}, \bibinfo {author} {\bibfnamefont {O.~B.}\ \bibnamefont {Korneta}},
  \bibinfo {author} {\bibfnamefont {S.}~\bibnamefont {Parkin}}, \bibinfo
  {author} {\bibfnamefont {J.}~\bibnamefont {Hu}}, \ and\ \bibinfo {author}
  {\bibfnamefont {G.}~\bibnamefont {Cao}},\ }\href {\doibase
  10.1103/PhysRevB.85.165143} {\bibfield  {journal} {\bibinfo  {journal} {Phys.
  Rev. B}\ }\textbf {\bibinfo {volume} {85}},\ \bibinfo {pages} {165143}
  (\bibinfo {year} {2012})}\BibitemShut {NoStop}%
\bibitem [{\citenamefont {Hu}\ \emph {et~al.}(2018)\citenamefont {Hu},
  \citenamefont {Shen}, \citenamefont {Hao}, \citenamefont {Liu}, \citenamefont
  {Wang}, \citenamefont {Sun},\ and\ \citenamefont
  {Shen}}]{10.3389/fchem.2018.00438}%
  \BibitemOpen
  \bibfield  {author} {\bibinfo {author} {\bibfnamefont {F.}~\bibnamefont
  {Hu}}, \bibinfo {author} {\bibfnamefont {F.}~\bibnamefont {Shen}}, \bibinfo
  {author} {\bibfnamefont {J.}~\bibnamefont {Hao}}, \bibinfo {author}
  {\bibfnamefont {Y.}~\bibnamefont {Liu}}, \bibinfo {author} {\bibfnamefont
  {J.}~\bibnamefont {Wang}}, \bibinfo {author} {\bibfnamefont {J.}~\bibnamefont
  {Sun}}, \ and\ \bibinfo {author} {\bibfnamefont {B.}~\bibnamefont {Shen}},\
  }\href {\doibase 10.3389/fchem.2018.00438} {\bibfield  {journal} {\bibinfo
  {journal} {Frontiers in Chemistry}\ }\textbf {\bibinfo {volume} {6}},\
  \bibinfo {pages} {438} (\bibinfo {year} {2018})}\BibitemShut {NoStop}%
\bibitem [{\citenamefont {Pan}\ \emph {et~al.}(2017)\citenamefont {Pan},
  \citenamefont {Chen}, \citenamefont {Jiang}, \citenamefont {Hu},
  \citenamefont {Yu}, \citenamefont {Yamamoto}, \citenamefont {Ogata},
  \citenamefont {Hattori}, \citenamefont {Guo}, \citenamefont {Fan} \emph
  {et~al.}}]{pan2017colossal}%
  \BibitemOpen
  \bibfield  {author} {\bibinfo {author} {\bibfnamefont {Z.}~\bibnamefont
  {Pan}}, \bibinfo {author} {\bibfnamefont {J.}~\bibnamefont {Chen}}, \bibinfo
  {author} {\bibfnamefont {X.}~\bibnamefont {Jiang}}, \bibinfo {author}
  {\bibfnamefont {L.}~\bibnamefont {Hu}}, \bibinfo {author} {\bibfnamefont
  {R.}~\bibnamefont {Yu}}, \bibinfo {author} {\bibfnamefont {H.}~\bibnamefont
  {Yamamoto}}, \bibinfo {author} {\bibfnamefont {T.}~\bibnamefont {Ogata}},
  \bibinfo {author} {\bibfnamefont {Y.}~\bibnamefont {Hattori}}, \bibinfo
  {author} {\bibfnamefont {F.}~\bibnamefont {Guo}}, \bibinfo {author}
  {\bibfnamefont {X.}~\bibnamefont {Fan}},  \emph {et~al.},\ }\href {\doibase
  10.1021/jacs.7b08625} {\bibfield  {journal} {\bibinfo  {journal} {Journal of
  the American Chemical Society}\ }\textbf {\bibinfo {volume} {139}},\ \bibinfo
  {pages} {14865} (\bibinfo {year} {2017})}\BibitemShut {NoStop}%
\bibitem [{\citenamefont {Hamada}\ and\ \citenamefont
  {Takenaka}(2011)}]{doi:10.1063/1.3540604}%
  \BibitemOpen
  \bibfield  {author} {\bibinfo {author} {\bibfnamefont {T.}~\bibnamefont
  {Hamada}}\ and\ \bibinfo {author} {\bibfnamefont {K.}~\bibnamefont
  {Takenaka}},\ }\href {\doibase 10.1063/1.3540604} {\bibfield  {journal}
  {\bibinfo  {journal} {Journal of Applied Physics}\ }\textbf {\bibinfo
  {volume} {109}},\ \bibinfo {pages} {07E309} (\bibinfo {year}
  {2011})}\BibitemShut {NoStop}%
\bibitem [{\citenamefont {Saha}\ \emph {et~al.}(2017)\citenamefont {Saha},
  \citenamefont {Fauth}, \citenamefont {Caignaert},\ and\ \citenamefont
  {Sundaresan}}]{184107}%
  \BibitemOpen
  \bibfield  {author} {\bibinfo {author} {\bibfnamefont {R.}~\bibnamefont
  {Saha}}, \bibinfo {author} {\bibfnamefont {F.}~\bibnamefont {Fauth}},
  \bibinfo {author} {\bibfnamefont {V.}~\bibnamefont {Caignaert}}, \ and\
  \bibinfo {author} {\bibfnamefont {A.}~\bibnamefont {Sundaresan}},\ }\href
  {\doibase 10.1103/PhysRevB.95.184107} {\bibfield  {journal} {\bibinfo
  {journal} {Phys. Rev. B}\ }\textbf {\bibinfo {volume} {95}},\ \bibinfo
  {pages} {184107} (\bibinfo {year} {2017})}\BibitemShut {NoStop}%
\bibitem [{\citenamefont {Sage}\ \emph {et~al.}(2007)\citenamefont {Sage},
  \citenamefont {Blake}, \citenamefont {Marquina},\ and\ \citenamefont
  {Palstra}}]{195102}%
  \BibitemOpen
  \bibfield  {author} {\bibinfo {author} {\bibfnamefont {M.~H.}\ \bibnamefont
  {Sage}}, \bibinfo {author} {\bibfnamefont {G.~R.}\ \bibnamefont {Blake}},
  \bibinfo {author} {\bibfnamefont {C.}~\bibnamefont {Marquina}}, \ and\
  \bibinfo {author} {\bibfnamefont {T.~T.~M.}\ \bibnamefont {Palstra}},\ }\href
  {\doibase 10.1103/PhysRevB.76.195102} {\bibfield  {journal} {\bibinfo
  {journal} {Phys. Rev. B}\ }\textbf {\bibinfo {volume} {76}},\ \bibinfo
  {pages} {195102} (\bibinfo {year} {2007})}\BibitemShut {NoStop}%
\bibitem [{\citenamefont {Hemberger}\ \emph {et~al.}(2006)\citenamefont
  {Hemberger}, \citenamefont {Rudolf}, \citenamefont {Krug~von Nidda},
  \citenamefont {Mayr}, \citenamefont {Pimenov}, \citenamefont {Tsurkan},\ and\
  \citenamefont {Loidl}}]{PhysRevLett.97.087204}%
  \BibitemOpen
  \bibfield  {author} {\bibinfo {author} {\bibfnamefont {J.}~\bibnamefont
  {Hemberger}}, \bibinfo {author} {\bibfnamefont {T.}~\bibnamefont {Rudolf}},
  \bibinfo {author} {\bibfnamefont {H.-A.}\ \bibnamefont {Krug~von Nidda}},
  \bibinfo {author} {\bibfnamefont {F.}~\bibnamefont {Mayr}}, \bibinfo {author}
  {\bibfnamefont {A.}~\bibnamefont {Pimenov}}, \bibinfo {author} {\bibfnamefont
  {V.}~\bibnamefont {Tsurkan}}, \ and\ \bibinfo {author} {\bibfnamefont
  {A.}~\bibnamefont {Loidl}},\ }\href {\doibase 10.1103/PhysRevLett.97.087204}
  {\bibfield  {journal} {\bibinfo  {journal} {Phys. Rev. Lett.}\ }\textbf
  {\bibinfo {volume} {97}},\ \bibinfo {pages} {087204} (\bibinfo {year}
  {2006})}\BibitemShut {NoStop}%
\end{thebibliography}%

\end{document}